\documentclass[journal,10pt]{IEEEtran}

\usepackage{cite}
\usepackage{amsmath,amssymb,amsfonts}
\usepackage{algorithm}
\usepackage{algpseudocode}
\usepackage{graphicx}

\usepackage[caption=false,font=footnotesize]{subfig}
\usepackage{adjustbox}
\usepackage{textcomp}
\usepackage{xcolor}
\usepackage{hyperref}
\usepackage{xcolor}
\usepackage{optidef}
\usepackage{stfloats}
\usepackage{multirow}

\begin{document}
\bstctlcite{BSTcontrol}

\title{Standard-Compliant Neuromorphic Integrated  Sensing and Communications Aided by an Intelligent Reflecting Surface}

\author{Jiho~Park,~\IEEEmembership{Graduate Student Member,~IEEE}, Jiechen~Chen,~\IEEEmembership{Member,~IEEE}, Joonhyuk~Kang,~\IEEEmembership{Member,~IEEE}, and Osvaldo~Simeone,~\IEEEmembership{Fellow,~IEEE}  
\thanks{This work was partly supported by the Institute of Information \& Communications Technology Planning \& Evaluation (IITP)-ITRC (Information Technology Research Center) grant funded by the Korea government (MSIT) (IITP-2026-RS-2020-II201787, contribution rate: 50\%); in part by the Institute of Information \& Communications Technology Planning \& Evaluation (IITP) under 6G·Cloud Research and Education Open Hub grant funded by the Korea government (MSIT) (IITP-2026-RS-2024-00428780, contribution rate: 50\%). \textit{(Corresponding authors: Joonhyuk Kang and Osvaldo Simeone.)}\\
\indent J. Park and J. Kang are with the Department of Electrical Engineering, Korea Advanced Institute of Science and Technology, Daejeon 34141, South Korea (email: phw5150@kaist.ac.kr; jhkang@ee.kaist.ac.kr). \\
\indent J. Chen is with the Department of Engineering, King’s College London, London, WC2R 2LS, UK (email: jiechen.chen@kcl.ac.uk). \\
\indent O. Simeone is with the Institute for Intelligent Networked Systems, Northeastern University London, One Portsoken Street, London, E1 8PH, UK (email: o.simeone@northeastern.edu).
 }
\vspace*{-0.3cm}
}

\maketitle
\IEEEpeerreviewmaketitle

\vspace{-1cm}
\begin{abstract}
Neuromorphic computing enables event-driven, low-power inference and is therefore an attractive technology for jointly carrying out communication, sensing, and processing at resource-constrained wireless devices. Impulse radio ultra-wideband (IR-UWB) signaling, which is standardized in the IEEE 802.15.4 family and widely deployed for ranging and short-range communication, is naturally matched to neuromorphic processing, since both operate through sparse temporal events, or spikes. In this context, this paper studies a standard-compliant neuromorphic integrated sensing and communications (N-ISAC) system in which a single spiking neural network (SNN) receiver jointly demodulates digital data and detects a passive radar target directly from a common IEEE 802.15.4z high-rate-pulse-repetition-frequency (HRP) UWB waveform. Unlike prior N-ISAC work, which assumed a basic pulse-position-modulation interface and a simplified multipath channel, we consider the full standardized physical layer together with a realistic ray-traced channel aided by a reconfigurable intelligent surface (RIS). The model accounts for the frequency-selective response of a metamaterial-based RIS, which disperses the wideband IR-UWB pulses and can degrade both communication and sensing. Numerical experiments quantify the impact of RIS frequency selectivity on UWB ISAC and characterize the trade-off between ISAC performance and receiver computation energy.
\end{abstract}

\begin{IEEEkeywords}
Neuromorphic computing, spiking neural networks, integrated sensing and communications, intelligent reflecting surface.
\end{IEEEkeywords}

\section{Introduction}
\label{sec:introduction}
Neuromorphic computing is an emerging brain-inspired technology that processes information through sparse, asynchronous spikes rather than dense clocked arithmetic. Implemented via spiking neural networks (SNNs), neuromorphic processors carry out computation only when input events occur, enabling event-driven inference with low latency and low energy consumption at resource-constrained edge devices \cite{snn_srm_ref, computation_energy_ref}. This efficiency has motivated the integration of neuromorphic computing with Impulse radio ultra-wideband (IR-UWB)  transmission, giving rise to \emph{neuromorphic communications}, in which sparse IR pulses are processed natively by an SNN for low-power remote inference and semantic communication in wireless Internet-of-Things (IoT) networks \cite{snn_comm_ref, neurocom_ref}. The appeal of this approach stems from a structural match between IR signaling and neuromorphic processing, as both operate through sparse temporal events and thereby support efficient event-driven inference \cite{snn_comm_ref, nisac_ref, chen2024neuromorphic}.


\begin{figure}[htp]
    \centering
    \includegraphics[width=3.3in]{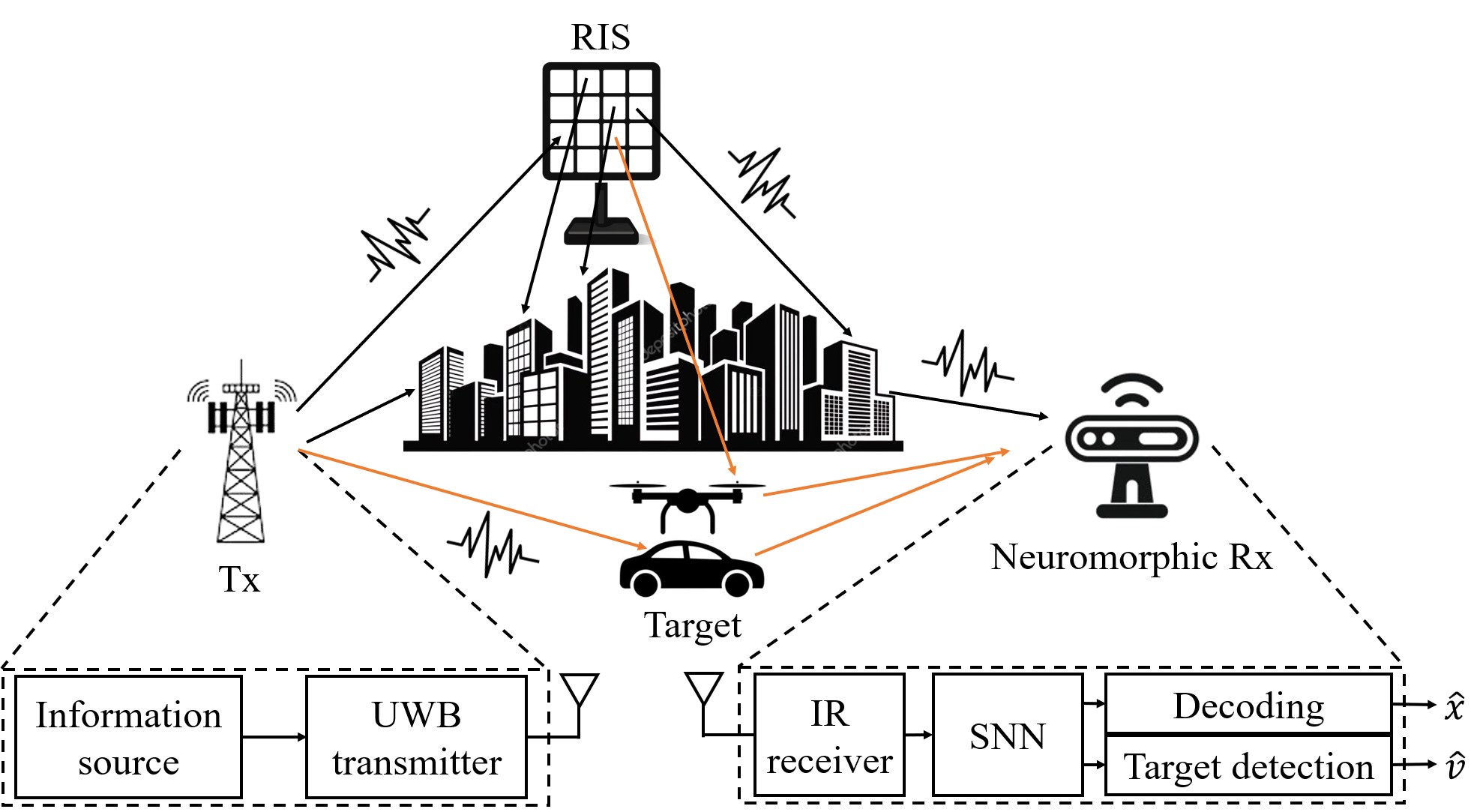}
    \caption[System model]{Illustration of the neuromorphic ISAC (N-ISAC) system under study in which a neuromorphic receiver equipped with a spiking neural network (SNN) performs simultaneous data demodulation and target detection. The system implements a standard-compliant UWB radio interface (see Fig.~\ref{fig2:tx_signal_block}), and the channel is modeled via ray tracing (RT) while accounting for the presence of a RIS.
    } \label{fig1:system_model}
\end{figure}

IR-UWB signaling also provides a suitable waveform for integrated sensing and communication (ISAC) because of its high temporal resolution,  low power consumption, and low implementation cost.  In addition to their original use in radar sensing systems, IR-UWB signals have been extensively studied for communication applications \cite{intro_ir_start_ref}. In particular, communication and ranging functionalities have been standardized within the IEEE 802.15.4 family of low-rate wireless networks  \cite{intro_ir_standreview1_ref, intro_ir_standreview2_ref, uwb_standz_ref}. IEEE 802.15.4z-compliant secure fine-ranging radios are now embedded in mainstream smartphones, automotive digital keys, and consumer tags for accurate localization and access control, making standard-compliant UWB transceivers a high-volume, low-power platform on which neuromorphic ISAC could be deployed at scale. Most recently, the IEEE 802.15.4ab amendment has reflected the emerging trend of incorporating sensing capabilities into conventional communication and ranging applications \cite{intro_ir_standab_ref}. 

Reconfigurable intelligent surfaces (RISs) have in parallel emerged as a key enabler for next-generation communication and sensing, since they can shape the propagation environment by establishing controllable non-line-of-sight (NLoS) links, improving coverage, and enhancing target illumination for ISAC  \cite{intro_ris_lim_ref}. However,  the frequency-dependent response of the metamaterial used in the design of the RIS must be taken into account for wideband transmissions such as IR-UWB  \cite{Lorentzian_ref}. Away from the resonance frequency, the RIS imposes amplitude attenuation and phase dispersion that broaden and distort the received pulses, potentially degrading both the time-of-arrival resolution required for sensing and the demodulation of the data symbols. Hence, while an RIS can be beneficial by creating additional reflected paths, its inherent dispersion may be detrimental to standard-compliant UWB ISAC and must be explicitly modeled. 

To preserve electromagnetic consistency, ray-tracing-based RIS implementations have been adopted to capture NLoS propagation with large deflection angles in urban environments \cite{intro_ris_rt1_ref, intro_ris_rt2_ref}. In particular, the macroscopic reradiation models in \cite{intro_ris_form1_ref, intro_ris_form2_ref} have been integrated into the open-source Sionna Ray Tracing (RT) framework to realize a physically consistent RIS model in building-canyon environments, in which the RIS is not treated as a simple reflecting object \cite{sionnart_ref}.

Motivated by these observations, the goal of this paper is to study a standard-compliant neuromorphic ISAC system operating over a realistic RIS-aided channel. As illustrated in Fig. \ref{fig1:system_model}, we adopt the full IEEE 802.15.4z high-rate-pulse-repetition-frequency (HRP) UWB physical layer at the transmitter, a ray-traced ISAC channel that includes a frequency-selective metamaterial RIS, and an SNN receiver that jointly performs data demodulation and target detection while explicitly controlling its spike-driven computation energy.

\vspace{-0.15in}
\subsection{Related Work}

The N-ISAC concept was introduced in \cite{nisac_ref}, where a common IR waveform and a single SNN receiver were shown to support efficient online data decoding and radar target detection, revealing synergies and trade-offs between the two functionalities. That work, however, assumed a basic pulse-position-modulation (PPM) interface and a simplified statistical multipath channel, and did not consider standardized signaling or RIS-aided propagation. More broadly, SNN-based receivers have been developed for digital communication, including spiking demodulation and detection \cite{snn_comm_ref}, spiking belief-propagation decoding of short channel codes \cite{snn_decode_ref}, and spiking receiver architectures derived from neural counterparts \cite{neuromorphicrx_ref}, while event-driven semantic communication and split inference have been explored for remote and low-power applications \cite{neurocom_ref, spike_sparsity_ref}. These works confirm the energy advantages of neuromorphic processing but target communication or inference in isolation rather than standard-compliant joint sensing and communication.

A complementary line of research has investigated the decoupling of communication and sensing information in UWB-ISAC systems, where both data demodulation and target or environment sensing rely on accurate time-of-arrival (ToA) estimation of nanosecond-scale impulses. As a result, ToA information is coupled with the modulated data symbols in the delay or phase domains, depending on the UWB modulation format, such as PPM or binary phase-shift keying (BPSK). Decoupling strategies have been proposed using various methodologies \cite{rel_fde_ref, rel_diff1_ref, rel_diff2_ref}. For example, for PPM-based UWB-ISAC, simultaneous channel sensing and delay-Doppler sensing can be achieved via differential demodulation with soft-information enhancement \cite{rel_diff1_ref}.

On the propagation side, accurate RIS modeling for wideband systems has motivated frequency-selective element responses, such as the Lorentzian model that links the reflection coefficient to the resonant behavior of the metamaterial \cite{Lorentzian_ref}, as well as ray-traced and physically consistent reradiation models \cite{intro_ris_form1_ref, intro_ris_form2_ref, sionnart_ref}. In parallel, 3GPP has standardized ISAC channel descriptions that decompose the received signal into background and target-induced components and that concatenate target-illumination and target-receiver links with direction-dependent radar cross-section terms \cite{3gpp_isac_channel_ref, target_channel_ref, ris_target_interaction_ref}. 

Finally, channel-adaptive neural receivers based on hypernetworks, which generate the parameters of a second network from observed pilots, have been used to acquire channel state information and to adapt receivers to varying channels \cite{hypernet_ref, hypercsi_ref, neurocom_ref,zecchin_ICL}. To the best of our knowledge, none of these strands has been combined into a standard-compliant neuromorphic ISAC receiver operating over a frequency-selective RIS-aided channel.

\vspace{-0.15in}
\subsection{Contributions}

This paper studies a standard-compliant N-ISAC system in which a single SNN receiver jointly demodulates digital data and detects a passive radar target from an IEEE 802.15.4z HRP UWB waveform propagating over a ray-traced, RIS-aided ISAC channel. In contrast to the original N-ISAC formulation \cite{nisac_ref}, which assumed a basic PPM interface and a simplified multipath model, the present work addresses a fully standardized radio interface and a realistic dispersive propagation environment. The main contributions are summarized as follows.

\noindent $\bullet$ \textit{Standard-compliant neuromorphic ISAC.} We develop an N-ISAC system based on the complete IEEE 802.15.4z HRP UWB physical layer (see Fig. \ref{fig2:tx_signal_block}).  This replaces the basic PPM signaling and simplified channel of \cite{nisac_ref} with standardized framing, enabling neuromorphic ISAC to be assessed under a commercially relevant interface.

\noindent $\bullet$ \textit{Frequency-selective RIS-aided ISAC channel.} We adopt a ray-traced ISAC channel, aligned with 3GPP TR 38.901 ISAC modeling, that comprises background, dynamic-clutter, and target-induced components with both direct and RIS-assisted target-interaction paths. The model incorporates a frequency-selective metamaterial RIS through a resonance-normalized Lorentzian response. This model allows us to analyze how the resulting dispersion broadens the IR-UWB pulses and affects joint communication and sensing over the UWB band.

\noindent $\bullet$ \textit{Preamble-driven channel-adaptive SNN receiver.} Unlike \cite{nisac_ref}, which adopted a joint training approach \cite{zecchin_ICL},  we propose a neuromorphic receiver in which a hypernetwork generates per-frame neuron-wise scaling coefficients for the SNN backbone directly from the synchronized standardized SYNC preamble observations (see Fig. \ref{fig1:system_model}). This provides adaptation to current channel conditions without explicit channel estimation, inversion, or equalization. 

\noindent $\bullet$ \textit{Energy-aware training and computation-energy model.} We formulate an energy-aware training objective that jointly optimizes communication and sensing performance while a sparsity-promoting regularizer controls the spike-driven synaptic activity, exposing a tunable trade-off between ISAC performance and event-driven computation energy. 

\noindent $\bullet$ \textit{Comprehensive evaluation.} Through extensive experiments with a standard-compliant transmitter and a ray-traced RIS-aided channel, we demonstrate the gains of preamble-driven channel adaptation, characterize the impact of RIS frequency selectivity on UWB ISAC, study the role of the standardized payload parameters, and assess the energy efficiency of the proposed neuromorphic receiver.

\section{System Model}
\label{sec:system_model}
As shown in Fig.~\ref{fig1:system_model}, we study a neuromorphic integrated sensing and communication (N-ISAC) system \cite{nisac_ref}, which consists of a single-antenna impulse radio ultra-wideband (UWB) transmitter (Tx), a single-antenna receiver (Rx), and a passive target. The same impulse radio UWB waveform is used for \textit{digital communication} and \textit{radar sensing} of the passive target at the Rx, which carries out both tasks using a single SNN \cite{nisac_ref}. In contrast to the prior N-ISAC study in \cite{nisac_ref}, which introduced N-ISAC assuming a basic PPM scheme and a simplified multipath channel model, here we adopt a standard-compliant radio interface \cite{uwb_standz_ref} and we consider a realistic ray-traced channel model aided by a reconfigurable intelligent surface (RIS).

\vspace{-0.15in}
\subsection{UWB Transmission}
\label{subsec:hrp_uwb_transmission}
\begin{figure}[t]
    \centering
    \includegraphics[width=3.6in]{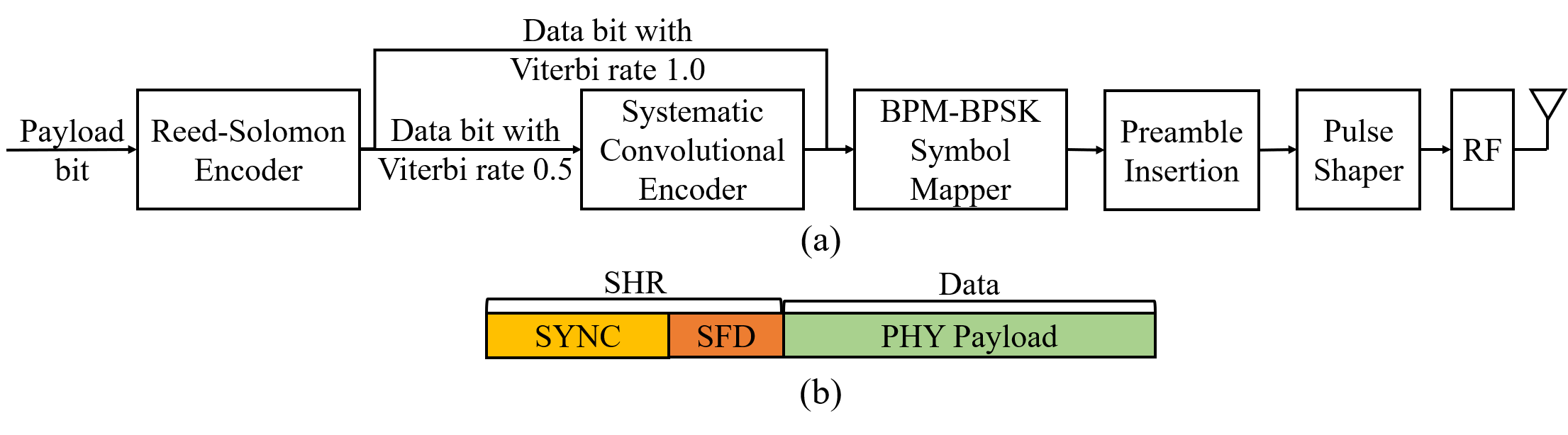}
    \vspace{-0.3in}
    \caption[HRP UWB frame]{High-rate pulse repetition frequency (HRP) UWB frame structure compliant with IEEE 802.15.4z \cite{uwb_standz_ref}. The information-bit sequence is encoded and mapped to burst-position modulation and binary phase-shift keying (BPM-BPSK) payload symbols, and each transmitted frame consists of a synchronization header (SHR) followed by a PHY payload.}
    \label{fig2:tx_signal_block}
\end{figure}

In the considered system, the Tx conveys a $D$-bit information sequence $\mathbf{x}=[x_0,x_1,\ldots,x_{D-1}]$, with $x_d\in\{0,1\}$, to the Rx during each transmitted frame. As illustrated in Fig.~\ref{fig2:tx_signal_block}, we adopt the IEEE 802.15.4z high-rate pulse repetition frequency (HRP) UWB PHY as the standard-compliant radio interface under study. The information bits are encoded through a forward error correction (FEC) chain and mapped to burst-position modulation (BPM) and binary phase-shift keying (BPSK) payload symbols. The resulting IR-UWB frame consists of a synchronization header (SHR) followed by a PHY payload: the SHR provides repeated known waveform observations for synchronization and frame-wise channel-dependent processing, while the PHY payload provides coded BPM-BPSK observations for joint data decoding and target detection.

Within the HRP UWB frame, the mean pulse repetition frequency (PRF) controls the temporal density of the transmitted impulses through standard-defined parameters. In the SHR, the mean PRF is set by the spreading length $\delta_p$, which determines the chip spacing of the known preamble code and hence the temporal structure of the preamble observation. In the PHY payload, the mean PRF determines the temporal density of the transmitted pulses and, together with the payload mode, determines the symbol duration and bit rate \cite{uwb_standz_ref}.

\begin{figure}[t]
    \centering
    \includegraphics[width=3.6in]{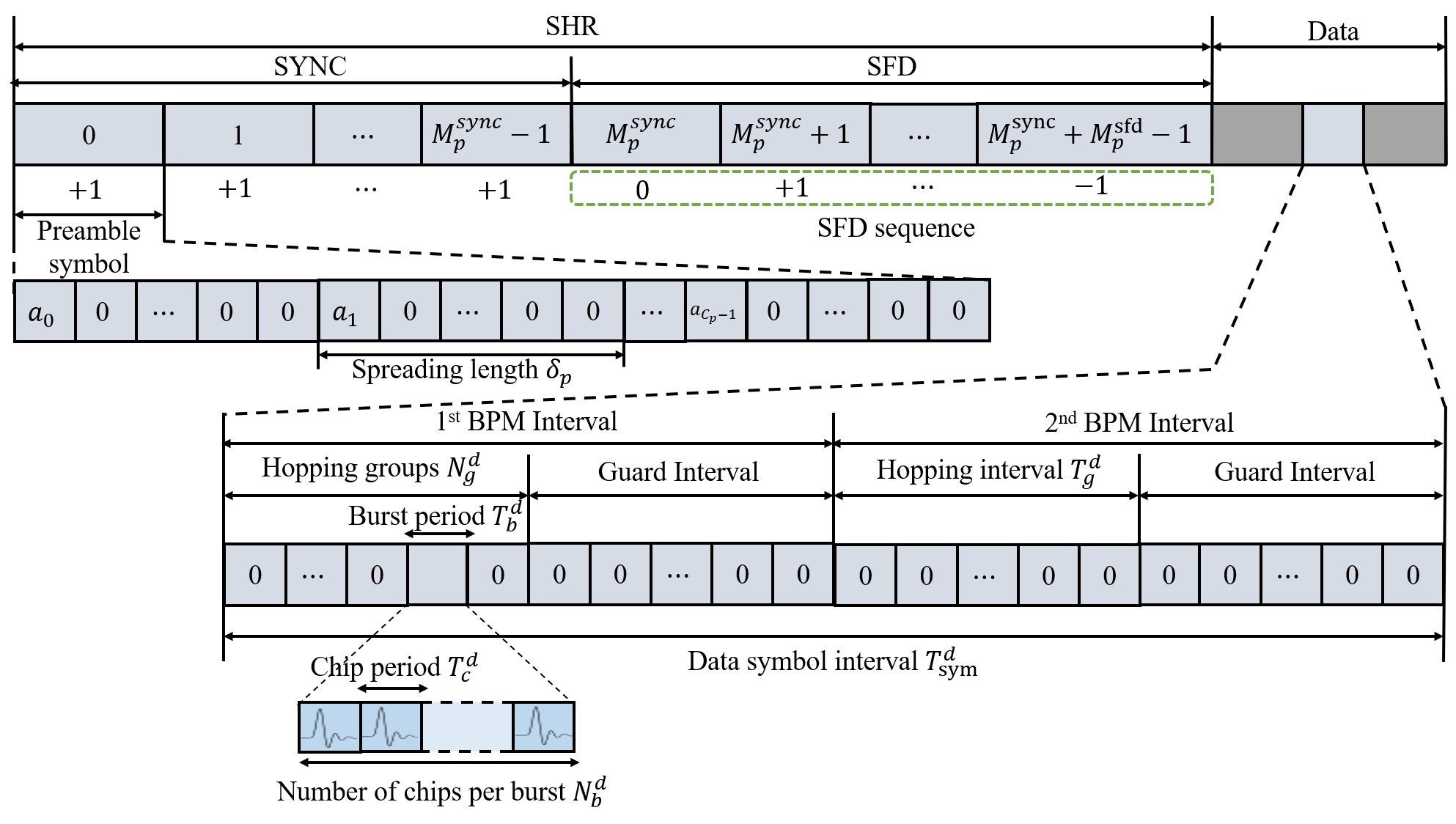}
    \caption[HRP UWB frame and symbol structure]{HRP UWB frame and symbol structure from IEEE 802.15.4z \cite{uwb_standz_ref}. The SHR contains SYNC and SFD preamble symbols generated from a spread ternary preamble code, while the PHY payload is composed of BPM-BPSK symbols. In each payload symbol, the BPM bit selects the active burst interval and the BPSK bit determines the burst polarity.}
    \label{fig3:frame_structure}
\end{figure}

As illustrated in Fig.~\ref{fig3:frame_structure}, the SHR is composed of a SYNC field and a start-of-frame delimiter (SFD), both of which are constructed from known preamble symbols. We define $M_p$ as the total number of SHR preamble symbols. Each preamble symbol is generated from a standard-defined ternary preamble code $\mathbf a=[a_0,\ldots,a_{C_p-1}]^T$, where $a_l\in\{-1,0,+1\}$ and $C_p$ is the preamble code length. The resulting preamble symbols are weighted by the polarity sequence $\mathbf e=[e_0,e_1,\ldots,e_{M_p-1}]^T$, where $e_m=1$ for $m=0,1,\ldots,M_p^{\mathrm{sync}}-1$, and $e_m\in\{-1,0,+1\}$ for $m=M_p^{\mathrm{sync}},\ldots,M_p-1$. The subsequence $\{e_m\}_{m=M_p^{\mathrm{sync}}}^{M_p-1}$ is referred to as the SFD sequence.
The preamble code $\mathbf{a}$ is spread by inserting $\delta_p-1$ zero-valued chips between consecutive ternary code entries, so that the total number of chips per preamble symbol is given by $N_c^p=C_p\delta_p$ \cite{uwb_standz_ref}. Accordingly, the transmitted SHR waveform is modeled as
\begin{equation}
    s_p(t) = \sum_{m=0}^{M_p-1} e_m \sum_{l=0}^{C_p-1}a_l p\left(t-mT_{\mathrm{sym}}^p-l\delta_p T_c\right),
    \label{eq:pre_sig_waveform}
\end{equation}
where $p(t)$ is the pulse waveform with bandwidth $B=1/T_c$, $T_c$ is the chip interval, and $T_{\mathrm{sym}}^p=N_c^pT_c$ is the preamble symbol duration.

For the PHY payload, the information sequence $\mathbf{x}$ is processed by the standard-aligned FEC chain shown in Fig.~\ref{fig2:tx_signal_block}. For payload modes with Viterbi rate $0.5$, the Reed-Solomon (RS) coded bits are passed through a convolutional encoder before BPM-BPSK mapping. For modes with Viterbi rate $1.0$, the convolutional encoder is bypassed, and the RS coded bits are directly mapped to BPM and BPSK bits.

Let $\mathbf{c}=[c_0,c_1,\ldots,c_{2M_d-1}]$ denote the coded payload-bit sequence produced by this FEC chain, where $M_d$ is the number of BPM-BPSK payload symbols. As illustrated in Fig.~\ref{fig3:frame_structure}, the two coded bits assigned to the $m$th BPM-BPSK symbol are given by
$d_m^{(0)} = c_{2m}$ and $d_m^{(1)} = c_{2m+1}$. The BPM bit $d_m^{(0)}\in\{0,1\}$ selects the active burst interval, whereas the BPSK bit $d_m^{(1)}\in\{0,1\}$ determines the burst polarity. Let $\zeta_{m,n}\in\{-1,+1\}$ denote the chip-sign coefficient applied to the $n$th chip of the active burst in the $m$th payload symbol, and let $\eta_m\in\{0,1,\ldots,N_g^d-1\}$ denote the corresponding burst-hopping address within the selected BPM interval. Accordingly, the transmitted payload waveform is modeled as \cite{uwb_standz_ref}
\vspace{-0.25in}
\begin{equation}
\begin{aligned}
s_d(t) &= \sum_{m=0}^{M_d-1}(-1)^{d_m^{(1)}}\sum_{n=0}^{N_b^d-1}\zeta_{m,n} \\
       &\quad\times p\left(t-mT_{\mathrm{sym}}^d - \left( 2d_m^{(0)}N_g^dN_b^d + \eta_mN_b^d+n \right)T_c \right),
\end{aligned}
\label{eq:pay_sym_waveform}
\end{equation}
where $T_{\mathrm{sym}}^d=N_c^dT_c$ is the payload symbol duration. The chip-sign sequence and burst-hopping address are generated according to the HRP UWB scrambling and burst-hopping rules. The burst-hopping sequence randomizes the active burst location within the selected BPM interval, whereas the chip-sign sequence controls the polarity of the chips in the active burst.

The pulse $p(t)$ is modeled using the standard-aligned impulse response of an eighth-order Butterworth filter with 3-dB bandwidth $B \simeq 500$ MHz\cite{uwb_standz_ref}. For each field $f\in\{p,d\}$, we impose the symbol-level energy constraint $E_s^f$, where $E_s^p$ and $E_s^d$ are the maximum symbol energies for the SHR preamble symbols and payload symbols, respectively.

\vspace{-0.4in}
\subsection{Frequency-Selective RIS-Aided Channel}
\label{subsec:channel_model}
\vspace{-0.1in}
The modulated waveform $s_f(t)$, with $f\in\{p,d\}$, is transmitted to the Rx through a RIS-assisted multipath channel. The channel depends on the possible presence of a radar target at a known delay cell. Let $v\in\{0,1\}$ denote the target state, where $v=0$ and $v=1$ correspond to the absence and presence of the target, respectively. Following the standard ISAC channel description in 3GPP TR 38.901 \cite{3gpp_isac_channel_ref}, the propagation channel is decomposed as
\vspace{-0.25in}
\begin{equation}
    h_v(t)=h_{\mathrm{bg}}(t)+v h_{\mathrm{tar}}(t),
    \label{eq:isac_channel}
\vspace{-0.1in}
\end{equation}
where $h_{\mathrm{bg}}(t)$ accounts for target-unrelated background clutter, and $h_{\mathrm{tar}}(t)$ accounts for target-induced echoes. 

To account for static and dynamic clutter in the multipath environment, we adopt a hybrid model for the background channel \cite{hybrid_channel_ref} that is implemented via RT. The static clutter is induced by fixed interaction objects, such as buildings, ground, and other stationary scatterers, whereas the dynamic clutter represents time-varying scattering due to moving objects. Accordingly, the background channel is modeled as
\begin{equation}
\begin{aligned}
h_{\mathrm{bg}}(t)
&=
\sqrt{\frac{K_{\mathrm{bg}}}{K_{\mathrm{bg}}+1}}
\sum_{p=1}^{P_{\mathrm{bg}}^{D}}
\alpha_{\mathrm{bg},p}^{D}
\delta\!\left(t-\tau_{\mathrm{bg},p}^{D}\right)
\\
&\quad+
\sqrt{\frac{1}{K_{\mathrm{bg}}+1}}
\sum_{p=1}^{P_{\mathrm{bg}}^{D}}
\sum_{b=1}^{B_{\mathrm{bg}}}
\alpha_{\mathrm{bg},p}^{D}\sqrt{w_{p,b}}\,
e^{j(\theta_{p,b}+2\pi f_c\Delta\tau_{p,b})}
\\
&\quad\times
\delta\!\left(t-\tau_{\mathrm{bg},p}^{D}
-\Delta\tau_{p,b}\right)
+
\sum_{n=1}^{N}
\alpha_{\mathrm{bg},n}^{R}
\psi\!\left(t-\tau_{\mathrm{bg},n}^{R}\right).
\end{aligned}
\label{eq:bg_channel}
\end{equation}
where the three terms, representing the static background component, the dynamic-clutter component, and the RIS-assisted component, respectively, are defined by the following quantities: the number of static paths $P_{\mathrm{bg}}^{D}$; the number of stochastic sub-rays generated from each static path $B_{\mathrm{bg}}$; and the number of RIS elements $N$; the complex amplitude and delay $\{\alpha_{\mathrm{bg},p}^{D}, \tau_{\mathrm{bg},p}^{D}\}$ of the $p$th static path; the corresponding pair $\{\alpha_{\mathrm{bg},n}^{R},\tau_{\mathrm{bg},n}^{R}\}$ for the RIS-assisted paths associated with the $n$th RIS element; the local excess-delay perturbation and random scattering phase $\Delta\tau_{p,b}$ and $\theta_{p,b}$ of the $b$th stochastic sub-ray generated from the $p$th non-RIS background path, respectively; the weights $w_{p,b}\ge0$, satisfying $\sum_{b=1}^{B_{\mathrm{bg}}}w_{p,b}=1$, that account for the power of the stochastic sub-rays; the deterministic-to-stochastic power ratio $K_{\mathrm{bg}}$ of the non-RIS background channel. The RIS response \(\psi(t)\) is described later in this section.

Following the 3GPP-aligned ISAC target-channel model \cite{target_channel_ref, ris_target_interaction_ref}, the target-induced channel $h_{\mathrm{tar}}(t)$ in \eqref{eq:isac_channel} is modeled through two target-interaction classes, namely the direct-incident class, denoted by $D$, and the RIS-incident class, denoted by $R$. For each path, the radar cross section (RCS) is decomposed into a mean term, a deterministic directional-pattern term, and a residual fluctuation term \cite{target_channel_ref}. Specifically, we write $\sigma_{p,q}^{D}=\sigma_{m}^{D}\sigma_{d,p,q}^{D}\sigma_{f}^{D}$ and $\sigma_{p,n}^{R}=\sigma_{m}^{R}\sigma_{d,p,n}^{R}\sigma_{f}^{R}$, where $\sigma_{m}^{b}$, $\sigma_{d,\cdot}^{b}$, and $\sigma_{f}^{b}$ denote the mean RCS level, direction-dependent term, and residual fluctuation term for $b\in\{D,R\}$, respectively. The path-pair-dependent directional term accounts for the incident and scattering directions at the target \cite{ris_target_interaction_ref}. The corresponding target-interaction factors are given by $\xi_{\mathrm{tar},p,q}^{D}=\sqrt{4\pi\sigma_{p,q}^{D}}/\lambda$ and $\xi_{\mathrm{tar},p,n}^{R}=\sqrt{4\pi\sigma_{p,n}^{R}}/\lambda$, where $\lambda$ is the carrier wavelength \cite{target_channel_ref}.

With these definitions, the target-induced channel is modeled as
\begin{equation}
\begin{aligned}
    h_{\mathrm{tar}}(t)
    &=
    \sum_{p=1}^{P_{\mathrm{out}}^{D}}
    \sum_{q=1}^{P_{\mathrm{in}}^{D}}
    \xi_{\mathrm{tar},p,q}^{D}
    \alpha_{\mathrm{out},p}^{D}
    \alpha_{\mathrm{in},q}^{D}
    \delta\left(t-\tau_{\mathrm{out},p}^{D}-\tau_{\mathrm{in},q}^{D}\right)
    \\
    &\quad+
    \sum_{p=1}^{P_{\mathrm{out}}^{D}}
    \sum_{n=1}^{N}
    \xi_{\mathrm{tar},p,n}^{R}
    \alpha_{\mathrm{out},p}^{D}
    \alpha_{\mathrm{in},n}^{R}
    \psi\left(t-\tau_{\mathrm{out},p}^{D}-\tau_{\mathrm{in},n}^{R}\right),
\end{aligned}
    \label{eq:target_channel}
\end{equation}
where the first term represents the direct-incident target echo, while the second term represents the RIS-incident target echo shaped by the dispersive RIS response. The latter is accounted for by the Lorentzian model \cite{Lorentzian_ref}
\begin{equation}
    \Psi(\omega) =  \left(\frac{\omega}{2\pi f_c}\right)^2\frac{j\kappa(2\pi f_c)}{(2\pi f_c)^2-\omega^2+j\kappa\omega},
    \label{eq:lorentzian_norm_freq}
\end{equation}
where $\kappa$ is the damping factor. The quality factor $Q\triangleq2\pi f_c/\kappa$ determines the bandwidth of the resonant response. Smaller values of $Q$ yield flatter frequency responses, while larger values of $Q$ yield narrower resonant profiles \cite{Lorentzian_ref}. The corresponding time-domain dispersive response is given by the inverse Fourier transform $\psi(t) =  \mathcal{F}^{-1}\{\Psi(\omega)\}$. In \eqref{eq:target_channel}, the symbols $P_{\mathrm{in}}^{D}$ and $P_{\mathrm{out}}^{D}$ denote the numbers of resolvable non-RIS incoming and outgoing target paths, respectively; the pair $\{\alpha_{\mathrm{in},q}^{D},\tau_{\mathrm{in},q}^{D}\}$ denotes the complex amplitude and delay of the $q$th non-RIS incoming Tx-target path; $\{\alpha_{\mathrm{in},n}^{R},\tau_{\mathrm{in},n}^{R}\}$ denotes the corresponding pair for the RIS-assisted incoming Tx-RIS-target path associated with the $n$th RIS element; the pair $\{\alpha_{\mathrm{out},p}^{D},\tau_{\mathrm{out},p}^{D}\}$ denotes the complex amplitude and delay of the $p$th outgoing target-Rx path. 

\vspace{-0.1in}
\subsection{Received Signal Processing}
For field $f\in\{p,d\}$, the signal obtained by the single-antenna Rx is given by
\begin{equation}
    y_f(t)=s_f(t)*h_v(t)+z_f(t),
    \label{eq:re_cont}
\end{equation}
where $s_f(t)$ is the transmitted waveform of field $f$; $h_v(t)$ is the effective ISAC channel in \eqref{eq:isac_channel}; $z_f(t)$ is additive white Gaussian noise with power spectral density $N_0$; and $*$ denotes convolution. The channel is assumed to remain constant over one transmitted frame.

Sampling signal \eqref{eq:re_cont} on the UWB chip grid at chip rate $1/T_c$ yields the discrete-time received sequence $y_{i,f}\triangleq y_f(iT_c)$ and the corresponding transmitted chip sequence $s_{i,f}\triangleq s_f(iT_c)$ for $i=0,1,\ldots,M_fN_c^f-1$, where $M_f$ and $N_c^f$ denote the number of symbols and the number of chips per symbol in field $f$, respectively. The samples in the $m$th symbol interval are collected as $\mathbf{y}_{m,f}=\{y_{i,f}\}_{i\in\mathcal{I}_m^f}$, where $\mathcal{I}_m^f\triangleq\{mN_c^f,\ldots,(m+1)N_c^f-1\}$.

For a maximum excess delay $T_h$, the effective channel has $L_h=\lceil T_h/T_c\rceil$ taps and is represented by $\mathbf{h}_v=[h_v(0),h_v(T_c),\ldots,h_v((L_h-1)T_c)]^T$. Accordingly, the $i$th chip-rate sample of field $f\in\{p,d\}$ is written as
\begin{equation}
    y_{i,f}=\mathbf{h}_v^T\mathbf{s}_{i,f}+z_{i,f},
    \label{eq:re_disc}
\end{equation}
where $\mathbf{s}_{i,f}=[s_{i,f},s_{i-1,f},\ldots,s_{i-L_h+1,f}]^T$ stacks the current and $L_h-1$ preceding transmitted chip samples, with the causal initialization $s_{i,f}=0$ for $i<0$. The additive noise sample is given by $z_{i,f}=z_f(iT_c)\sim\mathcal{CN}(0,N_0/T_c)$, where $N_0$ denotes the noise power spectral density.

For performance evaluation, we define the signal-to-noise ratio (SNR) after propagation through the channel. Since the AWGN sample in \eqref{eq:re_disc} has variance $N_0/T_c$, the average received per-chip SNR of field $f\in\{p,d\}$ is given by 
\begin{equation}
    \gamma_c^f\triangleq \frac{\mathbb{E}[|\mathbf{h}_v^T\mathbf{s}_{i,f}|^2]}{N_0/T_c},
    \label{eq:per_chip_snr}
\end{equation}
where the expectation is taken over the transmitted symbols, channel realizations, and target states used for evaluation.

In conventional coherent IR-UWB processing, the preamble timing is first acquired, the received SYNC samples are aligned with the known ternary preamble code, and the SFD is detected to determine the payload boundary \cite{synchronization_ref,sfd_detection_ref}. Following this Rx operation, we assume that the SHR timing and payload boundary are available at the Rx. Accordingly, the synchronized SYNC observations $\{\mathbf y_{m,p}\}_{m=0}^{M_p^{\mathrm{sync}}-1}$ and the payload observations $\{\mathbf y_{m,d}\}_{m=0}^{M_d-1}$ constitute the receiver-side interface for both the preamble-based coherent front end described in Section~\ref{sec:reference_frontend} and the neuromorphic Rx
introduced in Section~\ref{sec:proposed_receiver}.

\vspace{-0.1in}
\section{Preamble-Based Coherent UWB Front-End}
\label{sec:reference_frontend}

We next describe a preamble-based coherent UWB front end that uses the standardized SHR structure for channel-dependent receiver processing. In HRP UWB systems, the known SYNC field provides repeated preamble observations after timing acquisition and SFD detection, and these observations can be used to form correlation statistics, channel estimates, and equalized payload samples \cite{synchronization_ref,sfd_detection_ref,
conv_uwb_channel_estimation_ref}.

Since the SYNC field of the SHR waveform in \eqref{eq:pre_sig_waveform} consists of repeated unit-polarity preamble symbols, the waveform of one known SYNC symbol is written as
\begin{equation}
    r_p(t) = \sum_{l=0}^{C_p-1} a_l p\left(t-l\delta_p T_c\right),
\end{equation}
where $a_l\in\{-1,0,+1\}$ is the standard-defined ternary preamble code. Sampling this waveform on the chip grid gives
$r_{i,p}\triangleq r_p(iT_c)$ for $i=0,\ldots,N_c^p-1$. Since all SYNC symbols have unit polarity, the known SYNC chip sequence is
obtained by repeating the one-symbol template
$\mathbf{r}_p=[r_{0,p},r_{1,p},\ldots,r_{N_c^p-1,p}]^T$ over the
$M_p^{\mathrm{sync}}$ SYNC symbols. In particular, the transmitted chip in the $m$th SYNC symbol is given by
$s_{mN_c^p+i,p}=r_{i,p}$.

Let $S_p\triangleq M_p^{\mathrm{sync}}N_c^p$, and collect the synchronized SYNC samples as $\mathbf{y}_s=[y_{0,s},y_{1,s},\ldots,y_{S_p-1,s}]^T$. The transmission of the known SYNC sequence through the effective channel produces a channel-sensing observation that can be expressed in
convolution-matrix form as
\begin{equation}
    \mathbf{y}_s=\mathbf{P}_s\mathbf{h}_s+\mathbf{z}_s,
    \label{eq:re_channel_sensing}
\end{equation}
where $\mathbf h_s=[h_{0,s},h_{1,s},\ldots,h_{N_c^p-1,s}]^T$
denotes the effective chip-domain channel response represented over one preamble-symbol observation interval, and
$\mathbf{z}_s=[z_{0,s},z_{1,s},\ldots,z_{S_p-1,s}]^T$ is the
corresponding noise vector. The convolution matrix $\mathbf{P}_s\in\mathbb{C}^{S_p\times N_c^p}$ constructed from the complete known SYNC chip sequence uses entries $[\mathbf{P}_s]_{i,n}=s_{i-n,p}$, with $s_{i-n,p}=0$ whenever the chip index $i-n$ falls outside the SYNC field. Based on the channel-sensing observation in \eqref{eq:re_channel_sensing}, the correlation vector between the
synchronized SYNC samples and the known SYNC sequence is defined as $\mathbf{c}_s=\mathbf{P}_s^H\mathbf{y}_s$, where $(\cdot)^H$ denotes Hermitian transpose. Specifically, the $n$th element of $\mathbf{c}_s$ is given by $c_n^s=\sum_{i=0}^{S_p-1}s_{i-n,p}^{*}y_{i,s}$, where $(\cdot)^*$ denotes complex conjugation.

Following the conventional UWB channel estimators, we consider least-squares (LS) and statistical linear minimum mean-square error (LMMSE) channel estimation, with empirical channel mean and covariance estimated offline from the training channel realizations \cite{conv_uwb_channel_estimation_ref}. For the channel-estimation-based reference front ends, a causal payload convolution matrix $\hat{\mathbf{H}}_{d}\in\mathbb{C}^{N_c^d\times N_c^d}$ is constructed from either the LS or LMMSE channel estimate. Specifically, the matrix is formed from the estimated channel coefficients that contribute within one BPM-BPSK payload-symbol interval. We adopt the conventional MMSE equalizer for UWB communication systems \cite{conv_uwb_channel_estimation_ref}.

\vspace{-0.1in}
\section{Neuromorphic ISAC Receiver}
\label{sec:proposed_receiver}
We now describe the proposed neuromorphic Rx for the RIS-aided N-ISAC system described in Section~\ref{sec:system_model}. Building on the same standardized SHR observations used by the coherent front end in Section~\ref{sec:reference_frontend}, the proposed Rx replaces explicit channel estimation and payload equalization with preamble-driven hypernetwork adaptation. Prior to hypernetwork adaptation and payload processing, the synchronized SYNC and payload observations are sparsified in the chip domain as described in Section~\ref{subsec:input_encoding}. As illustrated in Fig.~\ref{fig1:system_model}, the Rx implements an SNN that adapts to the current channel condition via a hypernetwork operating on the sparsified SYNC observation. The channel-adapted SNN processes the sparsified BPM-BPSK payload observations and has task-specific readouts that produce coded-bit soft information for standard-aligned FEC decoding, as well as a frame-level target-presence indicator. In addition to optimizing communication and sensing performance, the proposed Rx design explicitly controls the sparsity of its spike-driven synaptic activity, enabling a tunable trade-off between joint ISAC performance and event-driven computation energy.

\vspace{-0.15in}
\subsection{Spiking Neural Network Backbone}
For the SNN backbone, we adopt the standard discrete-time spike response model (SRM) to describe the neuron dynamics \cite{snn_srm_ref}. To elaborate, let $\nu_{k,m}\in\{0,1\}$ denote the output spike of neuron $k$ at time step $m$, where $\nu_{k,m}=1$ represents spike emission and $\nu_{k,m}=0$ an idle state. Each neuron maintains a membrane potential $u_{k,m}$, which evolves as a function of the spikes received from presynaptic neurons and of its own past spikes. Accordingly, denoting $(\cdot)_m$ as the $m$th element of the argument signal, the membrane potential is modeled as \cite{neurocom_ref}
\begin{equation}
    u_{k,m} =  \sum_{j\in\mathcal{P}_k}w_{k,j}\left(\alpha * \nu_j\right)_m + \left(\beta * \nu_k\right)_m,
\end{equation}
where $\mathcal{P}_k$ is the set of presynaptic neurons connected to neuron $k$; $w_{k,j}$ is the synaptic weight from neuron $j$ to neuron $k$; and $*$ denotes convolution over the time index. The filter $\alpha$ denotes the synaptic response to incoming spikes, while the filter $\beta$ denotes the feedback response induced by the spike history of neuron $k$. Following the standard SRM description, the filter $\alpha$ accounts for temporal integration of presynaptic spikes, whereas the filter $\beta$ models the reset or refractory effect after spike emission \cite{snn_srm_ref}.

In the proposed implementation, we use a first-order leaky integrate-and-fire (LIF) realization of the SRM. Specifically, the membrane potential of hidden neuron $k$ is updated over the payload-symbol index $m$ as \cite{neuromorphicrx_ref}
\begin{equation}
    u_{k,m} = \eta_u u_{k,m-1} + \sum_{j\in\mathcal{P}_k} w_{k,j}\nu_{j,m} + b_k - \vartheta \nu_{k,m-1},
    \label{eq:lif_membrane}
\end{equation}
where $\eta_u\in[0,1)$ is the membrane decay factor; $w_{k,j}$ is the synaptic weight from presynaptic neuron $j$ to neuron $k$; $\nu_{j,m}$ is the spike emitted by presynaptic neuron $j$ at step $m$; $b_k$ is the bias current; and $\vartheta$ is the firing threshold. The first term in \eqref{eq:lif_membrane} implements the leaky memory of the membrane potential, the second and third terms define the instantaneous synaptic input current, and the last term implements the reset-by-subtraction induced by the previous spike of neuron $k$.

A spike is generated when the membrane potential crosses a threshold $\vartheta$, i.e.,
\begin{equation}
    \nu_{k,m} = \Theta(u_{k,m}-\vartheta),
    \label{eq:srm_spike}
\end{equation}
where $\Theta(\cdot)$ is the Heaviside step function. Accordingly, the hidden spiking neurons perform nonlinear event-driven processing while preserving temporal memory across successive received observations. Since synaptic computations are activated by presynaptic spike events, the computational activity of the SNN Rx depends on the spikes generated while processing the payload observations.

\vspace{-0.15in}
\subsection{Chip-Domain Sparse Input Encoding}
\label{subsec:input_encoding}

Preamble-driven adaptation and neural payload processing operate on continuous-valued chip-rate observations. Although the transmitted HRP UWB waveform is pulse based, the synchronized observations are generally dense owing to pulse shaping, multipath propagation, RIS-induced dispersion, and additive noise. We therefore sparsify the SYNC and payload observations prior to hypernetwork adaptation and neural inference. Magnitude-based sparsification suppresses
low-magnitude entries while retaining the dominant observations corresponding to the most relevant information about the timing of the transmitted pulses \cite{threshold_sparsification_ref}. Following ratio-controlled magnitude sparsification \cite{ratio_sparsification_ref}, the sparsification threshold is selected to retain a prescribed fraction of the largest-power chips.

Since the SYNC field consists of repeated unit-polarity preamble symbols, the corresponding observations are first coherently combined as
\begin{equation}
    \mathbf{y}_p = \sum_{m=0}^{M_p^{\mathrm{sync}}-1}\mathbf{y}_{m,p},
    \label{eq:raw_preamble_feature}
\end{equation}
where, for relative chip position $n=0,\ldots,N_c^p-1$, the $n$th entry of $\mathbf{y}_p$ is given by $[\mathbf{y}_p]_n = \sum_{m=0}^{M_p^{\mathrm{sync}}-1} y_{mN_c^p+n,p}$. Let $\rho\in(0,1]$ denote the prescribed fraction of retained chips. The SYNC sparsification threshold $\tau_p(\rho)$ denotes the $\lceil\rho N_c^p\rceil$th largest among the chip powers $|[\mathbf y_p]_n|^2$. The sparse SYNC observation at relative chip position $n$ is given by
\begin{equation}
    [\mathbf{y}_p^{s}]_n =
    \begin{cases}
        [\mathbf{y}_p]_n, & |[\mathbf{y}_p]_n|^2 \geq \tau_p(\rho), \\
        0, & \text{otherwise},
    \end{cases}
    \label{eq:sparse_preamble_input}
\end{equation}
where $\mathbf{y}_p^{s}\in\mathbb C^{N_c^p}$ retains the original complex values at the selected chip positions.

For the $m$th payload symbol, the sparsification threshold $\tau_{m,d}(\rho)$ is the $\lceil\rho N_c^d\rceil$th largest among the $N_c^d$ received chip powers $|y_{i,d}|^2$ for $i\in\mathcal I_m^d$. The corresponding sparse sample at chip index $i\in\mathcal{I}_m^d$ is given by
\begin{equation}
    y_{i,d}^{s} =
    \begin{cases}
        y_{i,d}, & |y_{i,d}|^2 \geq \tau_{m,d}(\rho), \\
        0, & \text{otherwise},
    \end{cases}
    \label{eq:sparse_payload_input}
\end{equation}
where $\mathbf{y}_{m,d}^{s} = \{y_{i,d}^{s}\}_{i\in\mathcal{I}_m^d} \in \mathbb{C}^{N_c^d}$ collects the sparse chip observations corresponding to the $m$th payload symbol. The same retained fraction $\rho$ is used for the SYNC and payload observations, while the corresponding thresholds are evaluated separately for the combined SYNC observation and for each payload symbol. Setting $\rho=1$ recovers the unsparsified observations.

\vspace{-0.15in}
\subsection{Hypernetwork-Based Channel Adaptation}

In a manner similar to \cite{neurocom_ref,hypercsi_ref}, in order to adapt the SNN Rx to the frame-dependent effective ISAC channel in \eqref{eq:isac_channel}, we introduce a hypernetwork that controls the synaptic weights of the SNN Rx based on the sparsified SYNC observation. Hypernetworks are neural networks whose outputs determine the parameters or parameter transformations of another network \cite{hypernet_ref}.

We first construct a real-valued representation of the sparsified SYNC observation $\mathbf y_p^s$ in \eqref{eq:sparse_preamble_input}. The corresponding real-valued hypernetwork input is
\begin{equation}
    \bar{\mathbf{y}}_p^s = \left[ \Re\{\mathbf{y}_p^s\}^{T}, \Im\{\mathbf{y}_p^s\}^{T} \right]^{T} \in \mathbb{R}^{2N_c^p},
    \label{eq:hypernet_input}
\end{equation}
where $\Re\{\cdot\}$ and $\Im\{\cdot\}$ denote the element-wise real and imaginary parts, respectively.

Let $N_L^R$ denote the number of inter-layer connections in the SNN Rx, and let $N_\ell^R$ denote the number of units in its $\ell$th layer. For the connection from layer $\ell$ to layer $\ell+1$, let $\tilde{\mathbf{W}}_\ell^R \in \mathbb{R}^{N_{\ell+1}^R \times N_\ell^R}$ denote the corresponding trainable channel-independent template weight matrix, and collect these matrices as $\tilde{\mathbf{W}}^R = \{\tilde{\mathbf{W}}_\ell^R\}_{\ell=1}^{N_L^R}$. Following the neuron-wise adaptation approach in \cite{neurocom_ref}, the hypernetwork maps the real-valued representation $\bar{\mathbf y}_p^s$ to the collection of layer-wise scaling vectors according to
\begin{equation}
    \mathbf{w}^R = \mathcal{H}_{\boldsymbol{\Phi}_H} \left( \bar{\mathbf{y}}_p^s \right),
    \label{eq:hypernet_scaling_factor}
\end{equation}
where $\mathbf{w}^R=[(\mathbf{w}_1^R)^T,\ldots,(\mathbf{w}_{N_L^R}^R)^T]^T$ and $\mathbf{w}_\ell^R \in \mathbb{R}^{N_\ell^R}$ contains one frame-dependent scaling coefficient for each unit in the $\ell$th layer. The channel-adapted weight matrix for the connection from layer $\ell$ to layer $\ell+1$ is then given by
\begin{equation}
    \mathbf{W}_\ell^R = \tilde{\mathbf{W}}_\ell^R \cdot \operatorname{diag} \left( \mathbf{w}_\ell^R \right),
    \label{eq:hypernet_adaptation}
\end{equation}
where $\operatorname{diag}(\mathbf{w}_\ell^R) \in \mathbb{R}^{N_\ell^R\times N_\ell^R}$ is the diagonal matrix formed from $\mathbf{w}_\ell^R$. Accordingly, the coefficient $w_{\ell,k}^R$ jointly scales all outgoing weights from the $k$th unit in layer $\ell$ to the units in layer $\ell+1$. We collect the resulting channel-adapted weight matrices as $\mathbf{W}^R=\{\mathbf{W}_\ell^R\}_{\ell=1}^{N_L^R}$.

Since the effective channel remains constant over one frame, the layer-wise scaling vectors collected in $\mathbf w^R$ and the adapted weight matrices collected in $\mathbf{W}^R$ are generated once from the sparsified SYNC observation and used for all payload symbols in the frame. During training, the hypernetwork parameters $\boldsymbol{\Phi}_H$ and the channel-independent template matrices $\tilde{\mathbf{W}}^R$ are jointly optimized. During inference, these trainable parameters remain fixed, while $\mathbf{w}^R$ and $\mathbf{W}^R$ are generated separately for each received frame.

\subsection{Joint Neuromorphic ISAC Inference}
\label{subsec:joint_neuromorphic_isac_inference}
Following the neuromorphic ISAC receiver architecture in \cite{nisac_ref}, in which a common SNN jointly processes the received IR waveform for data decoding and target detection, the channel-adapted SNN Rx parameterized by $\mathbf W^R$ sequentially processes the sparsified BPM-BPSK payload observations to produce communication and sensing outputs. For the $m$th payload symbol, the Rx constructs the real-valued SNN input from the sparsified payload observation $\mathbf{y}_{m,d}^{s}$ in \eqref{eq:sparse_payload_input} as
\begin{equation}
    \bar{\mathbf{y}}_{m,d}^{s} = \left[ \Re\{\mathbf{y}_{m,d}^{s}\}^{T}, \Im\{\mathbf{y}_{m,d}^{s}\}^{T} \right]^{T} \in \mathbb{R}^{2N_c^d}
    \label{eq:snn_rx_input}
\end{equation}
where $\Re\{\cdot\}$ and $\Im\{\cdot\}$ denote the element-wise real and imaginary parts, respectively, and $m=0,\ldots,M_d-1$. The input sequence $\{\bar{\mathbf{y}}_{m,d}^s\}_{m=0}^{M_d-1}$ is provided to the SNN Rx in payload-symbol order, allowing its internal membrane and spike states to evolve as successive observations are received. For each payload symbol, the readout layer associated with the final adapted weight matrix $\mathbf{W}_{N_L^R}^R$ produces the readout membrane potentials $u_m^{(0)}$, $u_m^{(1)}$, and $u_m^{(v)}$. The first two membrane potentials are associated with the BPM and BPSK coded bits, respectively, whereas the third provides target-presence information. In the following, $q=0$ and $q=1$ index the BPM and BPSK communication outputs, respectively.

For communication, the readout membrane potential $u_m^{(q)}$ is converted into the coded-bit soft probability as $\hat{p}_m^{(q)} = \sigma ( u_m^{(q)} )$, where $\sigma(x)=(1+e^{-x})^{-1}$ is the sigmoid function, and $\hat{p}_m^{(q)}$ represents the estimated probability that the coded bit $d_m^{(q)}$ is equal to one. To provide soft information compatible with the standard-aligned FEC decoder, we adopt the sigmoid-logit relation used for soft-output receiver processing in \cite{neuromorphicrx_ref} and define the corresponding coded-bit log-likelihood ratio (LLR) as
\begin{equation}
    L_m^{(q)} = \log \left( \frac{\hat{p}_m^{(q)}}{1-\hat{p}_m^{(q)}} \right) = u_m^{(q)},
    \label{eq:comm_llr}
\end{equation}
where $L_m^{(0)}$ and $L_m^{(1)}$ denote the LLRs associated with the BPM and BPSK coded bits, respectively. Hence, the communication readout membrane potentials are directly used as coded-bit soft information without forming intermediate hard BPM or BPSK decisions. The symbol-wise LLRs are arranged according to the coded payload-bit order as $\mathbf{L}_c = [ L_0^{(0)}, L_0^{(1)}, \ldots, L_{M_d-1}^{(0)}, L_{M_d-1}^{(1)} ]^{T} \in \mathbb R^{2M_d}$ where $L_m^{(q)}$ corresponds to the coded payload bit $d_m^{(q)}=c_{2m+q}$. The decoded information-bit sequence is then obtained as 
\begin{equation}
    \hat{\mathbf{x}} = \mathcal{R}_{\mathrm{FEC}} \left( \mathbf{L}_c \right) \in \{0,1\}^{D},
    \label{eq:fec_decoding}
\end{equation}
where $\mathcal{R}_{\mathrm{FEC}}(\cdot)$ denotes the standard-aligned FEC decoder corresponding to the encoding chain described in Section~\ref{sec:system_model}. For the configurations with Viterbi rate $0.5$, the decoder performs soft-input convolutional decoding followed by RS decoding, whereas the convolutional decoding stage is bypassed for the rate $1.0$ configurations.

For sensing, the third readout membrane potential is converted into a symbol-wise target-presence probability as $\hat{p}_m^{(v)} = \sigma ( u_m^{(v)} )$, where $\hat{p}_m^{(v)}$ represents the target-presence evidence available after processing the $m$th payload symbol. Motivated by the symbol-wise sensing outputs in \cite{nisac_ref}, the probabilities obtained over the payload duration are averaged to form the frame-level detection statistic expressed as
\begin{equation}
    \hat{p}^{(v)} = \frac{1}{M_d} \sum_{m=0}^{M_d-1} \hat{p}_m^{(v)},
    \label{eq:sensing_statistic}
\end{equation}
where $\hat p^{(v)}\in[0,1]$ summarizes the target-presence evidence collected throughout the payload duration. The frame-level target decision is then given by
\begin{equation}
    \hat{v} =
    \begin{cases}
        1, & \hat p^{(v)}\geq \dfrac{1}{2},\\
        0, & \mathrm{otherwise}
    \end{cases}
\end{equation}
where $\hat v=1$ and $\hat v=0$ indicate target presence and absence, respectively.

\vspace{-0.25in}
\subsection{Energy-Aware Training Problem}
\label{subsec:training_problem}
In this subsection, we formulate the offline supervised training of the proposed channel-adaptive SNN Rx by jointly accounting for communication and sensing performance and spike-driven synaptic activity. The set of trainable parameters is given by $\boldsymbol{\Theta} = \{ \tilde{\mathbf{W}}^{R}, \boldsymbol{\Phi}_H \}$, where $\tilde{\mathbf W}^{R}$ denotes the channel-independent template weights in \eqref{eq:hypernet_adaptation} and $\boldsymbol{\Phi}_H$ denotes the hypernetwork parameters in \eqref{eq:hypernet_scaling_factor}. We assume the availability of a training dataset $\mathcal{D} = \{ (\mathbf{x}_n,v_n) \}_{n=1}^{N_{\mathrm{train}}}$ containing $N_{\mathrm{train}}$ examples, where $\mathbf{x}_n \in \{0,1\}^{D}$ and $v_n\in \{0,1\}$ denote the information-bit sequence and target-presence label of the $n$th training example, respectively. For each $\mathbf{x}_n$, the standard-aligned FEC encoding and BPM-BPSK mapping described in Section~\ref{subsec:hrp_uwb_transmission} produce the coded-bit labels $d_{n,m}^{(q)}$, where $q=0$ and $q=1$ correspond to the BPM and BPSK bits of the $m$th payload symbol, respectively.

For each training pair $(\mathbf{x}_n,v_n)$, a realization of the effective channel in \eqref{eq:isac_channel}, conditioned on $v_n$, and the corresponding additive noise samples are drawn to generate the received frame according to \eqref{eq:re_disc}. The synchronized SYNC observations are coherently combined according to \eqref{eq:raw_preamble_feature}, sparsified according to \eqref{eq:sparse_preamble_input}, and converted into the real-valued hypernetwork input in \eqref{eq:hypernet_input}, while each payload observation is sparsified according to \eqref{eq:sparse_payload_input} and converted into the corresponding symbol-wise SNN input according to \eqref{eq:snn_rx_input}. The resulting communication and sensing probabilities $\hat p_{n,m}^{(q)}$ and $\hat p_{n,m}^{(v)}$ are obtained as described in Section~\ref{subsec:joint_neuromorphic_isac_inference}. Through the hypernetwork-based adaptation in \eqref{eq:hypernet_scaling_factor} and \eqref{eq:hypernet_adaptation}, these probabilities depend on the trainable parameters $\boldsymbol{\Theta}$, as well as on the sampled channel and noise realization.

Following the joint neuromorphic communication-and-sensing criterion in \cite{nisac_ref}, the communication performance for the $n$th training example is measured by the cross-entropy loss
\begin{equation}
\begin{aligned}
\mathcal{L}_{c}^{(n)}(\boldsymbol{\Theta})
&= -\frac{1}{2M_d}\sum_{m=0}^{M_d-1}\sum_{q=0}^{1}
\Bigl[d_{n,m}^{(q)}
\log \hat{p}_{n,m}^{(q)}(\boldsymbol{\Theta}) \\
&\qquad+
\left(1-d_{n,m}^{(q)}\right)
\log\left(1-\hat{p}_{n,m}^{(q)}(\boldsymbol{\Theta})\right)
\Bigr],
\end{aligned}
\end{equation}
where the loss is averaged over the BPM and BPSK coded bits of all payload symbols. Similarly, the sensing performance is measured by the cross-entropy loss
\begin{equation}
\begin{aligned}
\mathcal{L}_{s}^{(n)}(\boldsymbol{\Theta})
&= -\frac{1}{M_d}\sum_{m=0}^{M_d-1}
\Bigl[v_n\log \hat{p}_{n,m}^{(v)}(\boldsymbol{\Theta}) \\
&\qquad+
(1-v_n)
\log\left(1-\hat{p}_{n,m}^{(v)}(\boldsymbol{\Theta})\right)
\Bigr],
\end{aligned}
\end{equation}
where the same target-presence label $v_n$ supervises the sensing readout at every payload-symbol step, since the target state remains constant over the frame. The communication and sensing losses are combined into the joint ISAC training loss
\begin{equation}
    \mathcal{L}_{\mathrm{ISAC}}^{(n)}(\boldsymbol{\Theta}) = \chi \mathcal{L}_{c}^{(n)}(\boldsymbol{\Theta}) + (1-\chi) \mathcal{L}_{s}^{(n)}(\boldsymbol{\Theta}),
    \label{eq:isac_training_loss}
\end{equation}
where $\chi\in[0,1]$ determines the relative priority between communication and target detection.

Since spike-triggered synaptic operations constitute the activity-dependent computation of the SNN Rx, we introduce a sparsity-promoting regularizer to control the spiking activity while preserving joint communication-and-sensing performance. Following the membrane-potential sparsity regularization approach in \cite{spike_sparsity_ref}, let $u_{\ell,k,m}^{(n)}$ denote the membrane potential of the $k$th unit in hidden spiking layer $\ell$ when processing the $m$th payload symbol of the $n$th training example. For layer $\ell$ and payload-symbol step $m$, the corresponding sparsity regularizer is given by
\begin{equation}
    \mathcal{R}_{\ell,m}^{(n)}(\boldsymbol{\Theta}) = \frac{\left( \displaystyle \sum_{k=1}^{N_\ell^R} \left| \hat{u}_{\ell,k,m}^{(n)} \right| \right)^2}{\displaystyle \sum_{k=1}^{N_\ell^R} \left| \hat u_{\ell,k,m}^{(n)} \right|^2},
    \label{eq:spike_sparsity_regularizer}
\end{equation}
where $\hat{u}_{\ell,k,m}^{(n)} = \max\left(\frac{u_{\ell,k,m}^{(n)}}{\vartheta},0\right)$ denotes the threshold-normalized positive membrane potential. The ratio in \eqref{eq:spike_sparsity_regularizer} is scale-invariant and takes smaller values when the positive membrane-potential activity is concentrated in fewer units, thereby promoting sparse spiking activity.

Combining \eqref{eq:isac_training_loss} and \eqref{eq:spike_sparsity_regularizer}, the resulting energy-aware training objective is given by
\begin{equation}
    \min_{\boldsymbol{\Theta}} \sum_{n=1}^{N_{\mathrm{train}}} \mathbb{E}\left[ \mathcal{L}_{\mathrm{ISAC}}^{(n)}(\boldsymbol{\Theta}) + \frac{\lambda_r}{M_d} \sum_{m=0}^{M_d-1} \sum_{\ell\in\mathcal{S}_R} \mathcal{R}_{\ell,m}^{(n)}(\boldsymbol{\Theta}) \right],
    \label{eq:overall_train_objective}
\end{equation}
where $\mathcal{S}_R$ denotes the set of hidden spiking layers in the SNN Rx, regularization coefficient $\lambda_r\geq0$ controls the trade-off between neuromorphic ISAC performance and sparse spiking activity, and the expectation is taken over the target-dependent channel and additive noise realizations. In practice, the expectation in \eqref{eq:overall_train_objective} is approximated by sampling channel and noise realizations, and the resulting problem is addressed via stochastic gradient descent. To address the nondifferentiable Heaviside function in \eqref{eq:srm_spike}, we adopt the surrogate-gradient approach \cite{sgd_ref} by using the derivative of the differentiable unit-slope sigmoid function $\sigma(x)=(1+\exp(-x))^{-1}$ as a surrogate during backpropagation.

\section{Experiments}

In this section, we present numerical results to evaluate the performance and energy efficiency of the proposed standard-compliant N-ISAC system relative to reference receiver architectures. We consider a site-specific urban deployment in which an Rx-focused frequency-selective RIS assists a shadowed UWB link, while the same IEEE 802.15.4z HRP UWB waveform is used for data decoding and passive-target detection. The experiments characterize the impact of preamble-driven channel adaptation, RIS frequency selectivity, and standardized SHR and BPM-BPSK payload configurations, and assess the performance-energy trade-off enabled by chip-domain sparse input encoding, event-driven SNN processing, and spike-sparsity regularization.

\vspace{-0.1in}
\subsection{Experimental Setting}
\begin{figure}[htp]
    \centering
    \includegraphics[width=3.6in]{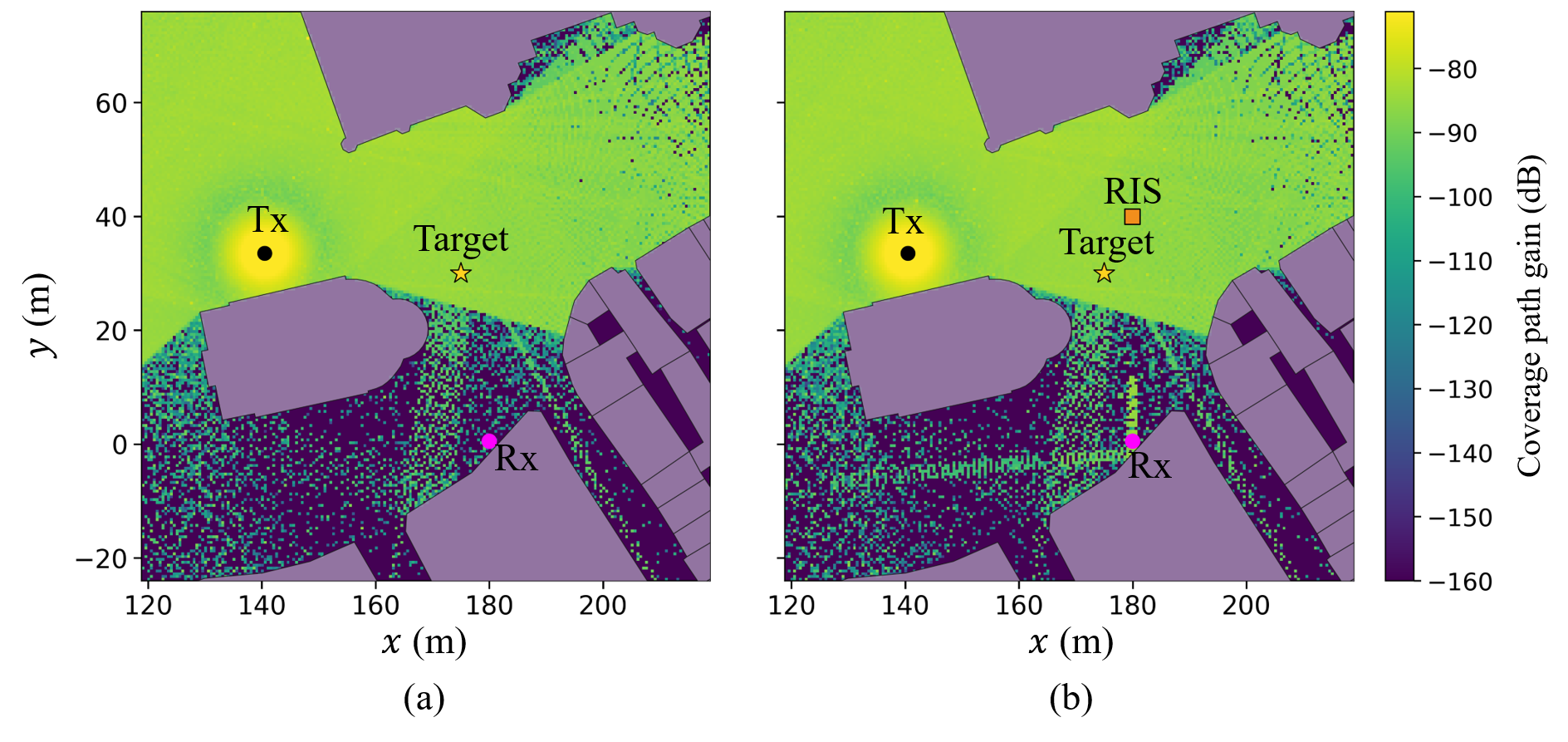}
    \vspace{-0.3in}
    \caption[Ray-traced urban deployment]{Site-specific path-gain maps generated using Sionna RT \cite{sionnart_ref}: (a) without the RIS and (b) with an Rx-focused RIS. The Tx, Rx, RIS, and target locations are indicated in both panels, which cover the same area and use a common color scale.
    } \label{fig5:sionna}
\end{figure}

We consider the site-specific urban propagation environment illustrated in Fig.~\ref{fig5:sionna}. The Tx, Rx, target center, and RIS center are located at $\mathbf{c}_{\mathrm{tx}}=(140.5,33.5,1.5)$ m, $\mathbf{c}_{\mathrm{rx}}=(180,0.5,1.5)$ m, $\mathbf{c}_{\mathrm{tar}}=(175,30,1)$ m, and $\mathbf{c}_{\mathrm{ris}}=(180,40,5)$ m, respectively. The Rx-focused RIS increases the path gain at the shadowed Rx from $-111.6$ dB to $-85.1$ dB, corresponding to a $26.5$ dB gain and enabling the evaluation of the communication-sensing trade-off induced by RIS-assisted propagation.

Using the OpenStreetMap-derived scene and Sionna RT \cite{sionnart_ref}, we generate $1200$ ray-traced channel realizations for training and $200$ disjoint realizations for testing. Each non-RIS path is augmented with $B_{\mathrm{bg}}=4$ stochastic sub-rays with $K_{\mathrm{bg}}=1$, while the target orientation and residual RCS fluctuation are resampled for each target-present frame, with the latter following a unit-mean lognormal distribution with $6$-dB standard deviation \cite{rcs_measurement_ref}. Each run uses $60\,000$ training frames and $10\,000$ test frames per SNR, with results averaged over five independent runs.

Unless stated otherwise, we adopt IEEE 802.15.4z HRP UWB channel~5 with $f_c=6.4896$ GHz and $B=499.2$ MHz \cite{uwb_standz_ref}. The SHR uses the high-mean-PRF configuration with $\delta_p=16$, $N_c^p=496$, $M_p^{\mathrm{sync}}=16$, and $M_p^{\mathrm{sfd}}=8$, while the BPM-BPSK payload uses $N_c^d=512$, $N_b^d=16$, and $N_g^d=8$, corresponding to $0.85$ Mb/s. We set $E_s^p=E_s^d=1$, $D=80$, and $v\sim\operatorname{Bern}(1/2)$. Unless varied, the RIS comprises $N=2500$ elements with quality factor $Q=20$, while the SNN and hypernetwork have $N_1^R=128$ and $1024$ hidden units, respectively.

To isolate the effects of receiver processing, the standard-compliant UWB waveform configuration, and RIS frequency selectivity from those of the sparsity mechanisms, all experiments preceding the energy-efficiency analysis use the unsparsified and unregularized setting, i.e., $\rho=1$ and $\lambda_r=0$. In the energy-efficiency analysis, we vary the input-retention ratio as $\rho\in\{0.5,0.7,0.9,1\}$ for the HNet SNN and ANN receivers, while fixing the regularization coefficient to $\lambda_r=10^{-4}$ for the SNN receivers. We further evaluate the effect of the coefficient $\lambda_r$ for HNet SNN at $\rho=0.7$ over $\{0, 10^{-5}, 5\times10^{-5}, 10^{-4}, 5\times10^{-4}, 10^{-3}\}$. Throughout, we set $\chi=0.5$ in \eqref{eq:isac_training_loss}.

To obtain a single Rx that operates over a range of channel conditions, we train each neural Rx once over the received per-chip SNR set $\Gamma_{\mathrm{train}}=\{-10,-5,0,5,10\}$ dB. Specifically, the SNR of each training frame is drawn independently and uniformly from $\Gamma_{\mathrm{train}}$, with $\gamma_c^p=\gamma_c^d=\gamma_c$, and the corresponding noise variances are set according to \eqref{eq:per_chip_snr} while $E_s^p$ and $E_s^d$ remain fixed. The same trained Rx is then evaluated at each test SNR without retraining.

\vspace{-0.15in}
\subsection{Baseline Schemes}

We evaluate the proposed \emph{HNet SNN} Rx against two classes of reference configurations designed to isolate the roles of preamble-driven adaptation and RIS-assisted propagation. The first class follows the coherent front end in
Section~\ref{sec:reference_frontend}, forming an explicit channel estimate and an equalized payload representation before neural inference. The second considers the corresponding no-RIS deployment while retaining the same transmitter and receiver-processing architecture. Unless otherwise stated, all SNN-based reference receivers in the following experiments use the same SNN backbone and communication and sensing readouts as the proposed Rx.

In detail, the baseline schemes are as follows:

\noindent $\bullet$ \textit{LS/LMMSE SNN:}
We consider two channel-estimation-based variants using LS and statistical LMMSE channel estimation, respectively, followed by MMSE payload equalization as described in Section~\ref{sec:reference_frontend}. The resulting equalized payload observations are processed by an SNN with parameters fixed across channel realizations. No hypernetwork-based adaptation is performed, and the chip-domain sparse input encoding in Section~\ref{subsec:input_encoding} is not applied.

\noindent $\bullet$ \textit{No-RIS:}
For each receiver architecture considered in the corresponding experiment, we also evaluate a no-RIS configuration in which the RIS-assisted background and target-interaction components in \eqref{eq:bg_channel} and \eqref{eq:target_channel}, respectively, are removed. The transmitter, receiver-processing architecture, standard-compliant waveform configuration, and evaluation procedure are otherwise unchanged. This reference configuration is used to isolate the impact of RIS-assisted propagation on communication, target detection, and receiver computation energy.

The proposed \emph{HNet SNN} Rx instead performs preamble-driven adaptation without explicit channel estimation or payload equalization. For energy-efficiency evaluation, we additionally consider ANN counterparts of HNet SNN and LS/LMMSE SNN, obtained by replacing the binary spiking activation with a continuous sigmoid activation while retaining the corresponding receiver architecture.

\vspace{-0.15in}
\subsection{Receiver Computation Energy Model}
\label{subsec:computation_energy}

We evaluate the computation energy required by each Rx to process one received HRP UWB frame using the operation-based model in \cite{neuromorphicrx_ref}. The total energy is computed as
\begin{equation}
    E_{\mathrm{Rx}} = \varepsilon_{\mathrm{MAC}}N_{\mathrm{MAC}} + \varepsilon_{\mathrm{AC}}N_{\mathrm{AC}},
    \label{eq:rx_computation_energy}
\end{equation}
where $N_{\mathrm{MAC}}$ and $N_{\mathrm{AC}}$ are the total numbers of multiply-accumulate (MAC) and accumulation (AC) operations, respectively, and we use $\varepsilon_{\mathrm{MAC}}=4.6$ pJ and $\varepsilon_{\mathrm{AC}}=0.9$ pJ for 32-bit floating-point operations in 45-nm CMOS \cite{computation_energy_ref}. The operation counts include receiver-dependent front-end processing, payload projection, and neural inference. For HNet, the front end comprises chip-domain sparse input encoding, preamble feature extraction, hypernetwork inference, and frame-wise adaptation, while LS and LMMSE include channel estimation and MMSE payload equalization. The ANN evaluates the payload projection densely, whereas the SNN skips zero-valued encoded inputs for $\rho<1$, using graded input events consistent with Loihi~2 \cite{loihi2_signal_ref,loihi2_spike_ref}.

For neural processing, synaptic operations triggered by binary hidden spikes are counted as AC operations, consistently with event-driven neuromorphic processing \cite{loihi1_ref}. The corresponding activity is characterized by the empirical firing rate of hidden layer $\ell$, defined as
\begin{equation}
    R_s(\ell) = \frac{1}{N_{\mathrm{test}}M_dN_\ell^R}\sum_{n=1}^{N_{\mathrm{test}}}\sum_{m=0}^{M_d-1}\sum_{k=1}^{N_\ell^R} \nu_{\ell,k,m}^{(n)},
    \label{eq:firing_rate}
\end{equation}
where $N_{\mathrm{test}}$ is the number of evaluated frames, $M_d$ is the number of payload symbols per frame, $N_\ell^R$ is the number of neurons in hidden layer $\ell$, and $\nu_{\ell,k,m}^{(n)}\in\{0,1\}$ denotes the spike emitted by neuron $k$ at payload-symbol step $m$ for test frame $n$. Accordingly, $R_s(\ell)$ gives the empirical probability of spike emission at a payload-symbol step, and the corresponding spike-dependent AC count scales as
$M_dR_s(\ell)N_\ell^R(N_{\ell+1}^R+1)$, where $N_{\ell+1}^R$ accounts for postsynaptic accumulations and the additional unit for the reset operation. The firing rate is evaluated separately for each SNN Rx, the input-retention ratio $\rho$, the sparsity regularization coefficient $\lambda_r$, and the RIS deployment. Training, channel-statistics precomputation, memory access, RF-front-end processing, RIS control and operation, and FEC decoding are excluded.

\vspace{-0.15in}
\subsection{Preamble-Driven Channel Adaptation}

\begin{figure}[t]
    \centering
    \includegraphics[width=2.85in]{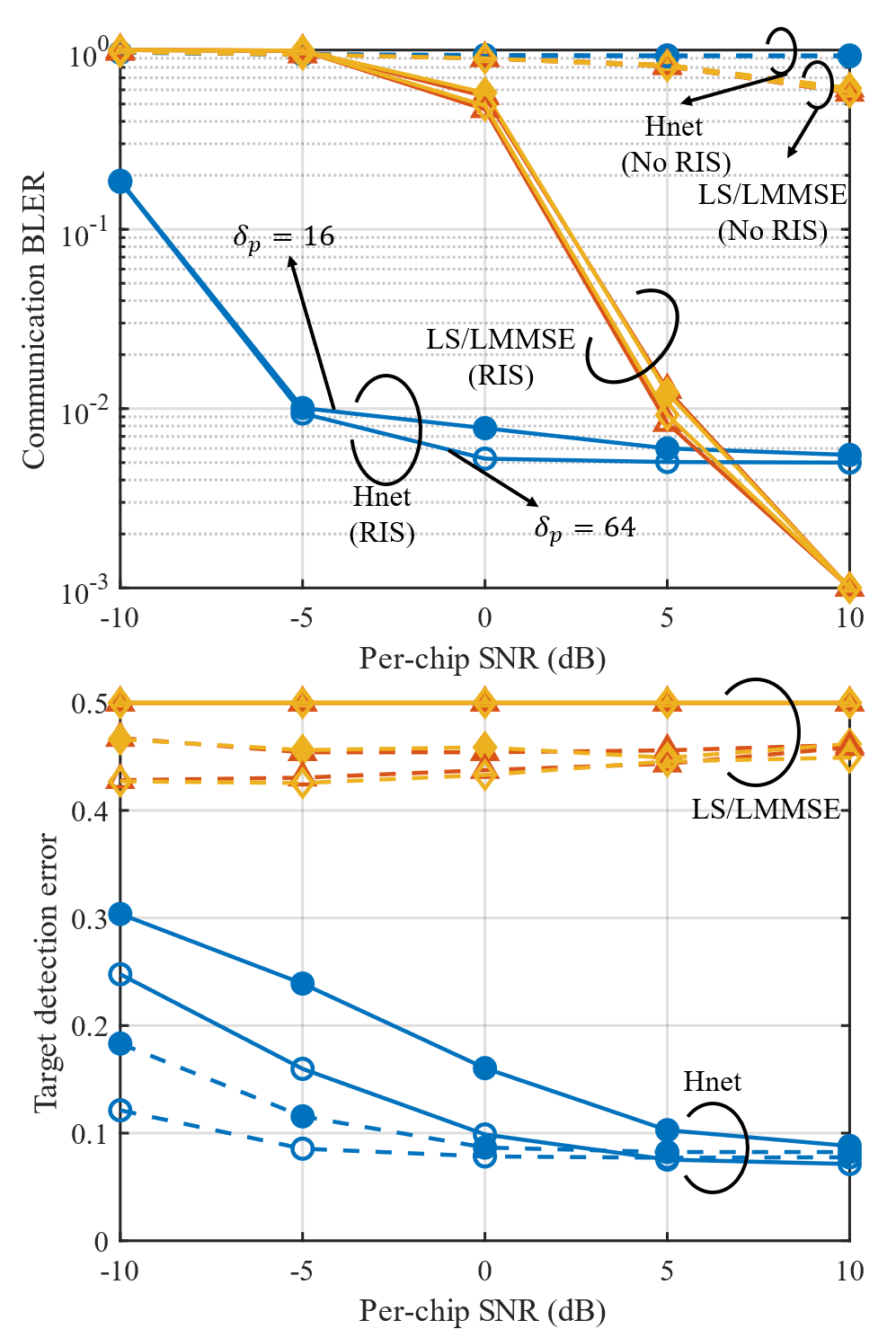}
    \vspace{-0.1in}
    \caption[Adaptation results]{(Top) Communication BLER and (Bottom) target-detection error versus received per-chip SNR $\gamma_c$ in \eqref{eq:per_chip_snr} for preamble-driven hypernetwork adaptation and channel-estimation-based equalized SNN Rxs across the RIS-aided and no-RIS deployments under the high- and low-mean-PRF SHR configurations; i.e., $\delta_p=16$ and $\delta_p=64$, respectively ($Q=20$, $N_c^d=512$, $N_b^d=16$, $\rho=1$, and $\lambda_r=0$).}
    \label{fig6:adaptation_result}
\end{figure}

We first investigate whether the IEEE 802.15.4z SYNC field can provide an effective interface for frame-wise task-oriented adaptation. Unlike prior hypernetwork-based wireless receivers relying on generic or jointly optimized pilots \cite{neurocom_ref,hypercsi_ref}, HNet SNN directly reuses the synchronized SYNC observations already available in the SHR, without additional pilot signaling. Fig.~\ref{fig6:adaptation_result} compares HNet SNN with channel-estimation-based LS and LMMSE SNN receivers under the high- and low-mean-PRF SHR configurations. The comparison is carried out for both the RIS-aided deployment and the corresponding no-RIS configuration.

As shown in Fig.~\ref{fig6:adaptation_result}, HNet SNN provides the most favorable joint communication and sensing performance, achieving a BLER of approximately $10^{-2}$ at per-chip SNR $\gamma_c=-5$ dB and reducing the target-detection error below $0.1$ at high SNR under the RIS-aided deployment. LS SNN and LMMSE SNN can outperform HNet SNN in terms of BLER at high SNR, but their target-detection errors remain close to the chance level of $0.5$. This result indicates that the equalized payload representation, although effective for data recovery, is not an effective input for the target-detection task in the considered setting. Comparing the RIS-aided and no-RIS deployments reveals a communication-sensing trade-off. The Rx-focused RIS substantially improves communication reliability for all considered receivers, but increases the target-detection error of HNet SNN at low and moderate SNR. Since the two deployments are compared at the same received per-chip SNR $\gamma_c$, this trade-off can be attributed to the RIS-induced channel structure: the strengthened target-independent background facilitates data recovery while masking target-dependent variations.

Comparing the high- and low-mean-PRF configurations, we note that the low-mean-PRF SHR generally improves both communication and sensing performance for HNet SNN by increasing the separation between nonzero preamble pulses, and hence reducing the overlap among multipath replicas. The more pronounced improvement under the RIS-aided deployment indicates that the increased pulse separation is particularly useful for mitigating the overlap among RIS-dispersed replicas. This improvement, however, increases the number of chips per preamble symbol $N_c^p$ from $496$ to $1984$ and the SYNC duration from $15.9~\mu$s to $63.6~\mu$s, thereby increasing both adaptation latency and preamble-processing cost.

\vspace{-0.15in}
\subsection{Impact of RIS Deployment}

\begin{figure}[t]
    \centering
    \includegraphics[width=2.85in]{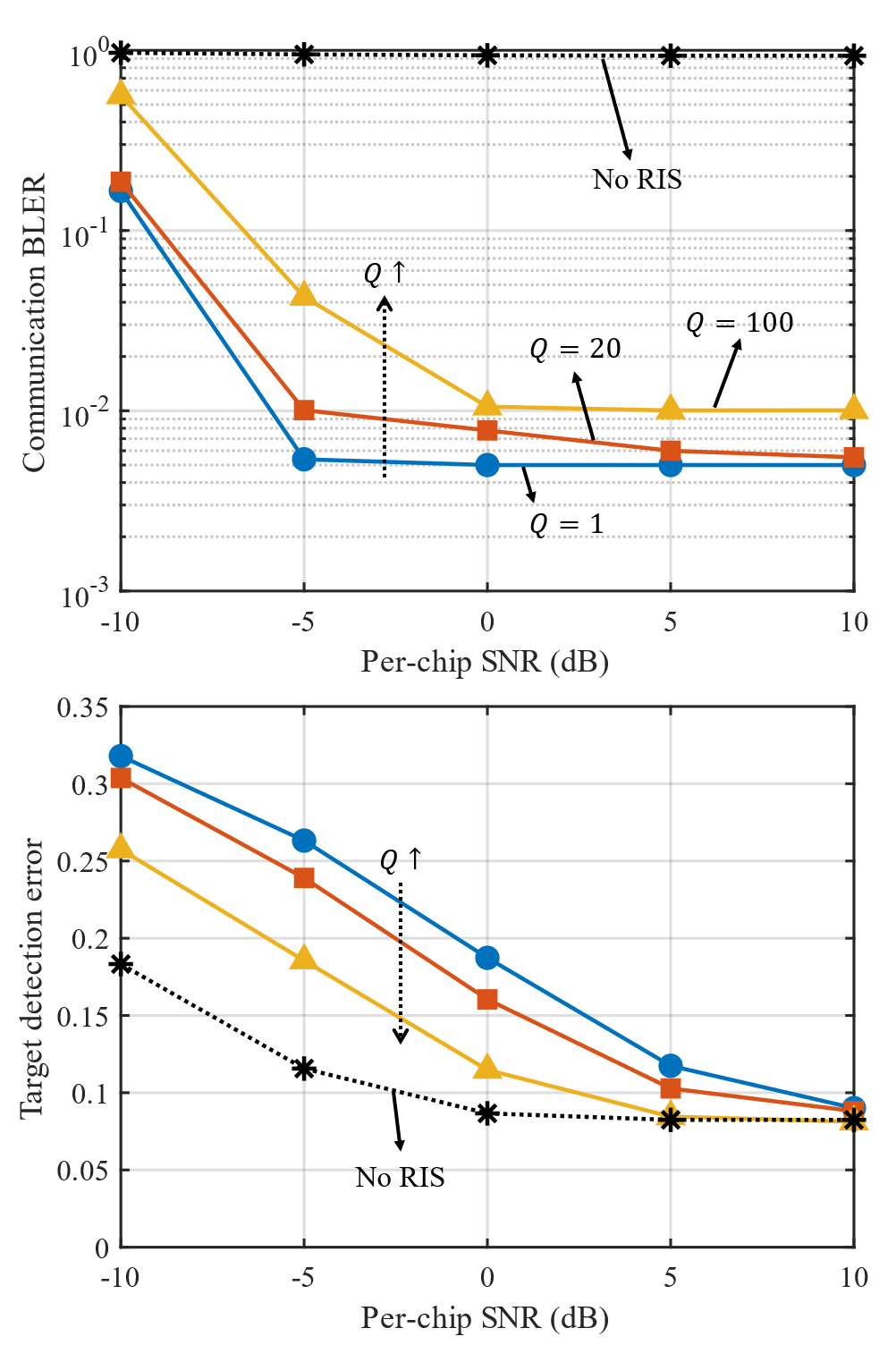}
    \vspace{-0.1in}
    \caption[Frequency selectivity trade-off results]{(Top) Communication BLER and (Bottom) target-detection error versus received per-chip SNR $\gamma_c$ for RIS quality factors $Q\in\{1,20,100\}$ ($\delta_p=16$, $N_c^d=512$, $N_b^d=16$, $\rho=1$, and $\lambda_r=0$).}
    \label{fig7:frequency_selectivity_result}
\end{figure}

Fig.~\ref{fig7:frequency_selectivity_result} shows the BLER and the target detection error as a function of per-chip SNR $\gamma_c$ for different values of the RIS quality factor $Q$. The figure shows that a larger RIS frequency selectivity induces opposite trends for communication and target detection. A low-$Q$ RIS provides a broadband response that better preserves the Rx-focused RIS-assisted component, yielding the lowest communication BLER. Moreover, as the quality factor $Q$ increases, off-resonant attenuation and temporal dispersion progressively degrade data decoding. Target detection instead improves with the RIS quality factor $Q$, since suppressing the target-unrelated RIS-assisted background makes the weaker target-dependent component more distinguishable and causes the detection error to approach that of the no-RIS case.

\begin{figure}[t]
    \centering
    \includegraphics[width=3in]{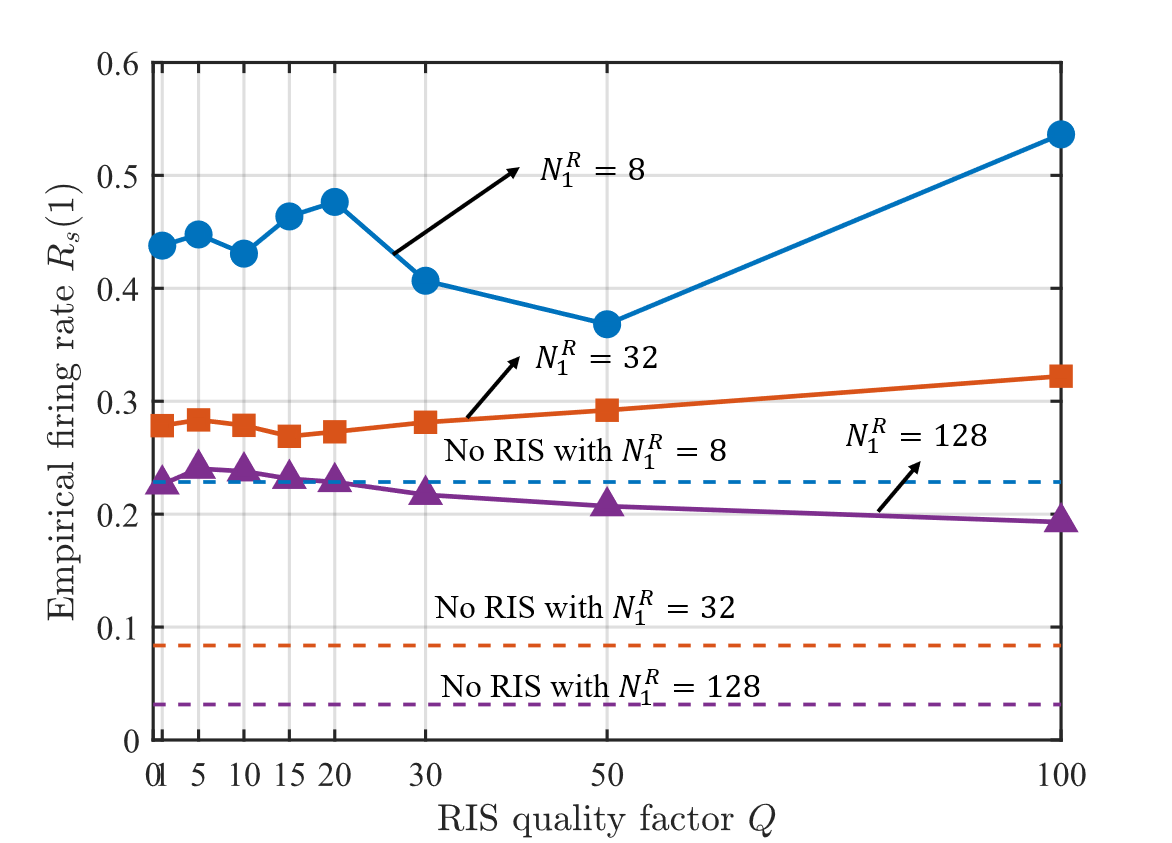}
    \vspace{-0.1in}
    \caption[Spike firing rate results]{Empirical firing rate $R_s(1)$ in \eqref{eq:firing_rate} versus RIS quality factor $Q$ for HNet SNNs with the number of hidden-layer units set to $N_1^R\in\{8,32,128\}$ ($\gamma_c=5$ dB, $\delta_p=16$, $N_c^d=512$, $N_b^d=16$, $\rho=1$, and $\lambda_r=0$).}
    \label{fig8:spike_firing_rate_result}
\end{figure}

Fig.~\ref{fig8:spike_firing_rate_result} shows that the RIS frequency selectivity also affects the energy consumption of the Rx through the empirical firing rate in \eqref{eq:firing_rate}. With $N_1^R=8$ hidden units, the SNN maintains a high and irregular firing rate with no systematic dependence on the quality factor $Q$. In contrast, wider SNNs exhibit lower firing rates and a more systematic variation with the quality factor $Q$. For example, with $N_1^R=128$, the firing rate decreases from approximately $0.24$ to $0.19$ as the RIS becomes strongly selective, following a modest increase at small $Q$. This behavior is consistent with the suppression of off-resonant RIS-assisted components, while the absence of a comparable trend for the narrow SNN indicates that the event-driven response to RIS selectivity is capacity dependent.

Overall, these results reveal a coupling between RIS frequency selectivity and neuromorphic processing. Increasing the quality factor $Q$ of the RIS improves target-detection performance at the expense of communication reliability, with a corresponding reduction in firing activity for sufficiently wide SNNs. The RIS response and neural capacity should therefore be co-designed to balance joint ISAC performance and event-driven computation.

\vspace{-0.15in}
\subsection{Standard-Compliant Payload Configuration}

\begin{figure}[t]
    \centering
    \includegraphics[width=2.85in]{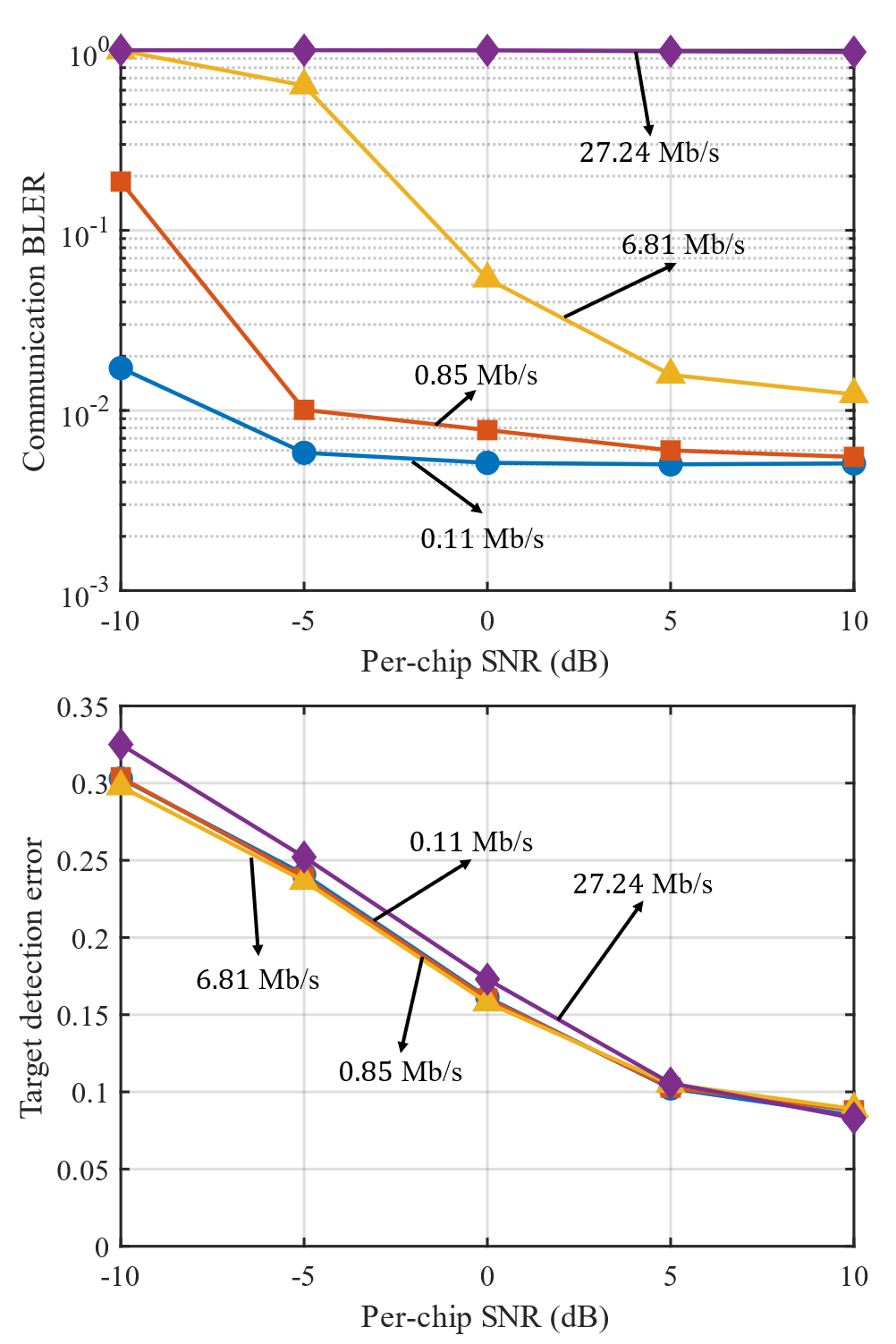}
    \vspace{-0.1in}
    \caption[Payload bit rate results]{(Top) Communication BLER and (Bottom) target-detection error versus received per-chip SNR $\gamma_c$ for the standard-compliant payload bit rates $\{0.11,0.85,6.81,27.24\}$ Mb/s, corresponding to $(N_b^d,N_c^d)\in \{(128,4096),(16,512),(2,64),(1,32)\}$ ($\delta_p=16$, $N_g^d=8$, $Q=20$, $\rho=1$, and $\lambda_r=0$).}
    \label{fig9:payload_bit_rate_result}
\end{figure}

Fig.~\ref{fig9:payload_bit_rate_result} evaluates the impact of high-mean-PRF payload modes of the HRP UWB PHY. For the considered modes corresponding to transmission rates $0.11$, $0.85$, and $6.81$ Mb/s, the Viterbi coding rate remains fixed at $1/2$, so that the performance variation is primarily due to a variation in symbol duration. As symbol duration decreases, and thus the rate increases, the ratio of the channel delay spread to the symbol duration grows, resulting in stronger overlap among delayed replicas of the dominant RIS-assisted component and a higher communication BLER. Over the same rate range, the target-detection error decreases moderately, showing that shorter symbols provide a modest sensing advantage. The $27.24$ Mb/s mode deviates from this trend because it bypasses convolutional coding. For this mode, severe temporal interference degrades both communication and sensing.

\vspace{-0.15in}
\subsection{Receiver Computation Energy}

\begin{figure}[t]
    \centering
    \includegraphics[width=2.85in]{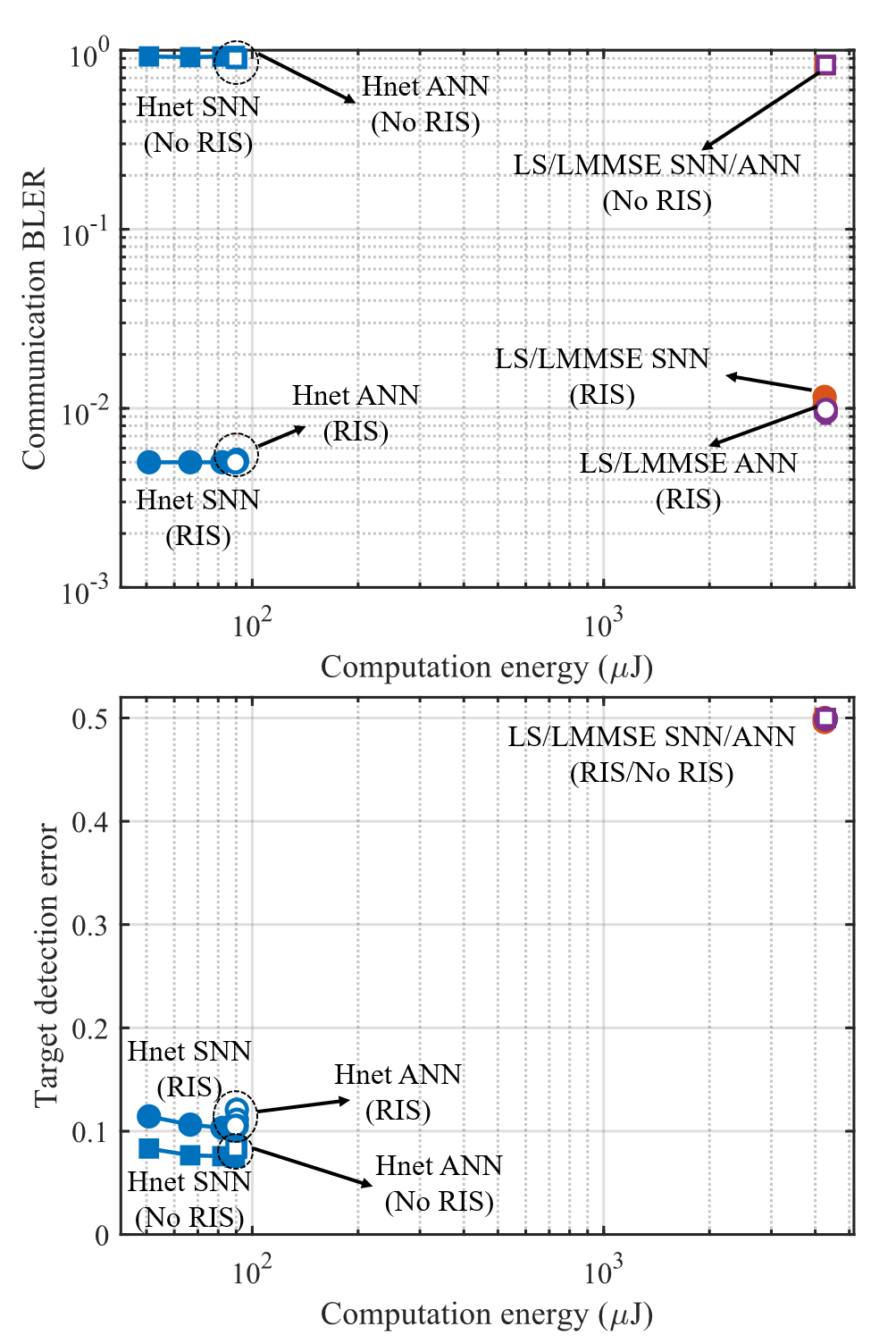}
    \vspace{-0.1in}
    \caption[Receiver energy performance results]{(Top) Communication BLER and (Bottom) target-detection error versus computation energy for HNet, LS, and LMMSE receivers with SNN and ANN backbones across the RIS-aided and no-RIS deployments. For HNet receivers, the curves are obtained by varying the input-retention ratio as $\rho\in\{0.5,0.7,0.9,1\}$ ($\gamma_c=5$ dB, $\lambda_r=10^{-4}$ for the SNN receivers, $Q=20$ for the RIS-aided deployment, $\delta_p=16$, $N_c^d=512$, and $N_b^d=16$).}
    \label{fig10:energy_performance_result}
\end{figure}

We now move on to evaluating computation energy and ISAC performance as a function of the Rx architecture. As shown in Fig.~\ref{fig10:energy_performance_result}, both the SNN and ANN implementations of the LS and LMMSE receivers require more than $4.2$ mJ per frame owing to explicit channel estimation and payload equalization, while providing no corresponding advantage in joint ISAC performance. By adapting directly from the synchronized chip-domain SYNC observations, the HNet receivers reduce the computation energy to below $91~\mu$J per frame, while attaining the most favorable joint operating region. 

Chip-domain sparse input encoding provides an additional energy reduction for HNet SNN. In particular, for the RIS-aided deployment, decreasing the input-retention ratio from $\rho=1$ to $\rho=0.5$ reduces the computation energy of HNet SNN from approximately $89~\mu$J to $51~\mu$J per frame, while maintaining a BLER of around $5\times10^{-3}$ and incurring only a moderate increase in target-detection error. In contrast, the computation energy of HNet ANN remains nearly unchanged with $\rho$, since its dense input projection does not exploit the zero-valued encoded samples. These results show that the dominant receiver-level energy savings are obtained by avoiding explicit channel estimation and payload equalization, while sparse input encoding further reduces the cost of event-driven SNN processing.

\begin{figure}[t]
    \centering
    \includegraphics[width=3in]{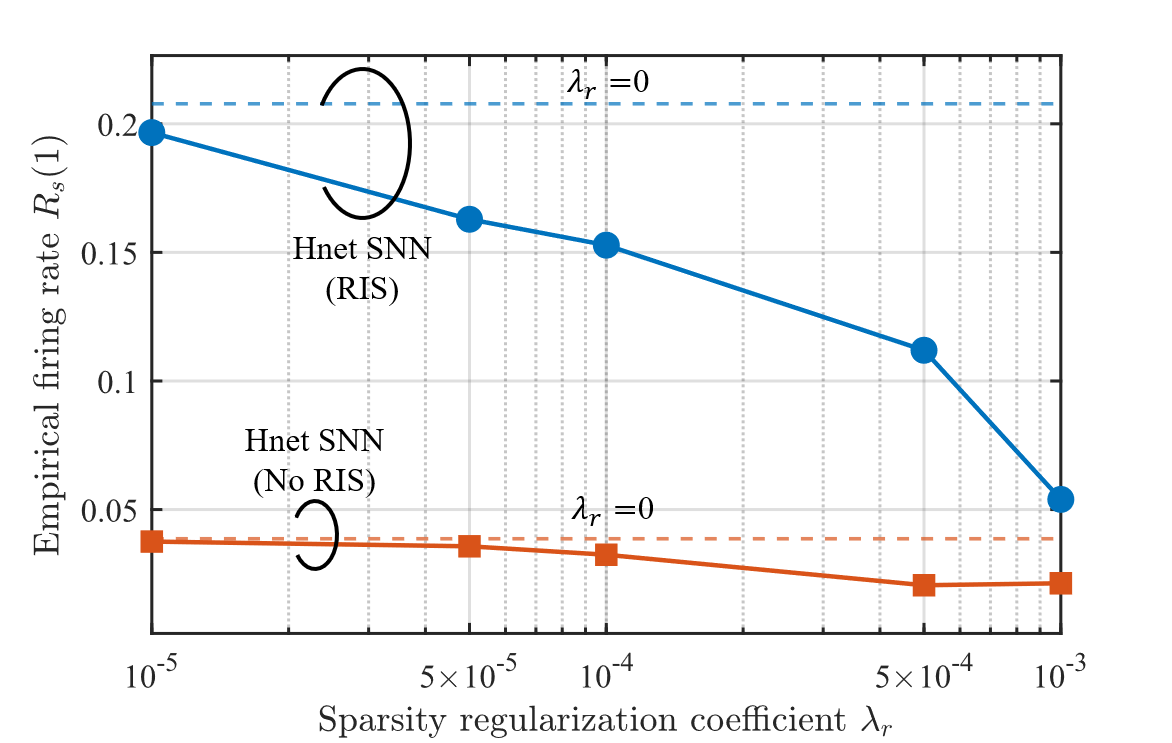}
    \vspace{-0.1in}
    \caption[Firing-rate regularization results]{Empirical firing rate $R_s(1)$ in \eqref{eq:firing_rate} versus the sparsity regularization coefficient $\lambda_r$ for HNet SNN across the RIS-aided and no-RIS deployments at $\rho=0.7$. The dashed horizontal lines correspond to the unregularized setting $\lambda_r=0$ ($\gamma_c=5$ dB, $Q=20$ for the RIS-aided deployment, $\delta_p=16$, $N_c^d=512$, and $N_b^d=16$).}
    \label{fig11:energy_firing_rate_result}
\end{figure}

The energy-efficiency gains enabled by sparse input encoding can be further complemented by controlling the spiking activity through the regularization coefficient $\lambda_r$. As shown in Fig.~\ref{fig11:energy_firing_rate_result}, for the RIS-aided deployment at the input-retention ratio $\rho=0.7$, increasing the coefficient $\lambda_r$ from $0$ to $10^{-4}$ reduces the empirical firing rate from approximately $0.21$ to $0.15$, with negligible variations in communication BLER and target-detection error. Increasing the coefficient $\lambda_r$ further to $10^{-3}$ reduces the firing rate to approximately $0.05$, corresponding to a reduction of about $74\%$ relative to the unregularized case, although a moderate degradation in communication reliability is observed. A similar but smaller reduction is observed for the no-RIS deployment, which exhibits lower firing rates across all values of the coefficient $\lambda_r$. Since the coefficient $\lambda_r$ affects only spike-dependent neural operations, its impact on the total Rx computation energy is smaller than that of chip-domain sparsification. Thus, the ratio $\rho$ and the coefficient $\lambda_r$ provide complementary means of controlling input sparsity and internal spiking activity, respectively.

\section{Conclusion}

This paper studied a standard-compliant N-ISAC system based on the IEEE 802.15.4z HRP UWB PHY over a ray-traced frequency-selective RIS-aided channel. The proposed HNet SNN adapts directly from standardized SYNC observations without additional pilots or explicit channel estimation and payload equalization, while chip-domain sparse input encoding and spike-sparsity regularization reduce receiver computation. The results reveal a communication-sensing trade-off induced by RIS frequency selectivity and show that the proposed receiver achieves favorable joint ISAC performance with substantially lower computation energy than channel-estimation-based baselines.

\bibliographystyle{IEEEtran}
\bibliography{reference}




\end{document}